\documentclass[12pt]{article}

\usepackage{amsmath,amssymb}
\usepackage{graphicx}
\usepackage{comment}

\newcommand{\mot}{\!\not \!}
\newcommand{\mmot}{\!\!\not \!\! }

\allowdisplaybreaks

\def\tr {{\rm Tr}}
\def\e {{\rm e}}
\def\c {\cos}
\def\s {\sin}
\def\d {{\rm d}}
\def\mot {\not \!}
\def\iy {\int_{-\pi R}^{\pi R}\!\!\!\!{\rm d}y}
\def\dx{\int_0^1\!\!{\rm d}x{\rm d}y~}                     
\def\E{\frac{\pi RM}{{\rm e}^{2\pi RM}-1}}     
\def\D{\mathcal{D}}                            
\def\nt{\notag}                                
\def\C{\mathcal{C}}                             
\def\dX{\int_0^1\!\!\!{\rm d}X}
\def\nt{\notag}
\def\dk{\int\!\!\frac{{\rm d}^4k}{(2\pi)^4i}}		
\def\dx{\int_0^1\!\!{\rm d}x{\rm d}y~}			
\def\e{{\rm e}}	
\def\E{\frac{\pi RM}{{\rm e}^{2\pi RM}-1}}	
\def\D{\mathcal{D}}				
				
\def\nt{\notag}					
\def\C{\mathcal{C}}		     	 
\def\dX{\int_0^1\!\!\!{\rm d}X}			
\def\gm{\gamma^{\mu}}

\makeatletter
    
    \@addtoreset{equation}{section}
\makeatother

\title{Lower Bound for Compactification Scale \\
\vspace*{2mm}
from Muon $g-2$ in the Gauge-Higgs Unification}
\author{Yuki Adachi, C.S. Lim and Nobuhito Maru${}^{\ast}$}
\date{}

\begin{document}
\setlength{\baselineskip}{18pt}
\begin{titlepage}

\begin{flushright}
KOBE-TH-09-01
\end{flushright}
\vspace{1.0cm}
\begin{center}
{\Large\bf Lower Bound for Compactification Scale \\ 
\vspace*{5mm}
from Muon $g-2$ in the Gauge-Higgs Unification} 
\end{center}
\vspace{25mm}

\begin{center}
{\large
Yuki Adachi, 
%
C. S. Lim
%
%
and Nobuhito Maru$^*$
}
\end{center}
\vspace{1cm}
\centerline{{\it Department of Physics, Kobe University,
Kobe 657-8501, Japan}}

\centerline{{\it
$^*$Department of Physics, Chuo University, 
Tokyo 112-8551, Japan
}}
%
%
\vspace{2cm}
\centerline{\large\bf Abstract}
\vspace{0.5cm}
We discuss a muon anomalous magnetic moment in a five dimensional 
$SU(3)$ gauge-Higgs unification compactified on $M^4\times S^1/Z_2$ space-time
including $Z_2$-odd bulk mass for a fermion.
We calculate one-loop corrections to the anomalous magnetic moment of muon 
and find a lower bound for the compactification scale to be around 5 $\sim$ 6 TeV
by comparing nonzero Kaluza-Klein mode contributions 
with the muon $(g-2)$ experiment at BNL. 
\end{titlepage}




\newpage

\section{Introduction}

Gauge-Higgs unification \cite{Manton, Fairlie, Hosotani} is one of the attractive scenarios 
solving the hierarchy problem without invoking supersymmetry. 
In this scenario, 
Higgs doublet in the Standard Model (SM) is identified with 
the extra spatial components of the higher dimensional gauge fields. 
A remarkable feature is that the quantum correction to Higgs mass is insensitive 
to the cutoff scale of the theory and calculable 
regardless of the non-renormalizability of higher dimensional gauge theory. 
The reason is that the Higgs mass term as a local operator is forbidden 
by the higher dimensional gauge invariance. 
The finite mass term is generated radiatively and expressed by the Wilson line phase as a non-local operator. 
This fact has opened up a new avenue to the solution of the hierarchy problem \cite{HIL}. 
Since then, much attention has been paid to the gauge-Higgs unification and 
many interesting works have been done from various points of view 
\cite{KLY}-\cite{LMgy}. 

The finiteness of Higgs mass has been studied and verified in various models 
and types of compactification at one-loop level\footnote{For the case of gravity-gauge-Higgs unification, 
see \cite{HLM}} \cite{ABQ}-\cite{LMH} and even at two loop level \cite{MY}.
It is natural to ask whether any other finite physical observables exist in the gauge-Higgs unification. 
The naive guess is that such observables are in the gauge-Higgs sector of the theory if they ever exist. 
Two of the present authors (C.S.L. and N.M.) studied the structure of divergences for S and T parameters 
in the gauge-Higgs unification since such parameters are described 
by higher dimensional gauge invariant operators with respect to gauge and Higgs fields, 
and are expected to be finite by virtue of the higher dimensional gauge symmetry. 
The result is that both parameters are divergent (convergent) more than (in) five dimensions 
as expected from the naive power counting argument.  
However, a nontrivial prediction we have found, specific to the gauge-Higgs unification, is 
that some linear combination of S and T parameters is finite even in six dimensions \cite{LM}. 

In a paper by the present authors \cite{ALM1}, 
we have found a more striking fact:  
we have shown that the magnetic moment of fermion 
in the $(D+1)$ dimensional QED gauge-Higgs unification model 
compactified on $S^1$ becomes finite for an arbitrary space-time dimension, 
regardless of the nonrenormalizability of the theory. 
Actually, the reason is very simple. 
In four dimensional space-time, 
a dimension six gauge invariant local operator describes the magnetic moment: 
\begin{align}
i \bar{\psi}_L \sigma^{\mu\nu} \psi_R F_{\mu\nu} \langle H \rangle. 
\label{MMM4}
\end{align}
However, when the operator is included into the scheme of gauge-Higgs unification, 
the Higgs doublet should be replaced 
by an extra space component of the higher dimensional gauge field $A_y$. 
Then the operator is forbidden by the higher dimensional gauge invariance, 
since $A_y$ transforms inhomogeneously under the gauge transformation. 
Then, to preserve the gauge symmetry, 
$A_y$ should be further replaced by gauge covariant derivative $D_y$, 
and the relevant gauge invariant operator becomes 
\begin{align}
i \bar{\Psi} \Gamma^{MN} D_L \Gamma^L \Psi F_{MN}  
\label{MMMD}
\end{align}
where $L, M$ and $N$ denote $D+1$ dimensional Lorentz indices. 
The key observation of our argument is that the operator (\ref{MMMD}), 
when $D_L$ is replaced by $\langle D_L \rangle$ with the gauge field $A_L$ replaced by its VEV, 
vanishes because of the on-shell condition $i \langle D_L \rangle \Gamma^L \Psi = 0$. 
From this fact, we can expect that the magnetic moment is finite 
and have shown that it is indeed the case by explicit diagrammatical calculations \cite{ALM1}. 
This is the specific prediction of the gauge-Higgs unification to be contrasted with 
the case of Randall-Sundrum model \cite{DHR} or the universal extra dimension scenario \cite{ADW, AD}, 
in which the magnetic moment of fermion diverges in the models with more than five space-time dimensions. 

Although this result was quite remarkable, the above model is too simple and not realistic. 
In particular, 
the gauge group $U(1)$ is too small to incorporate the standard model. 
In our previous paper \cite{ALM2}, 
we have studied more about the issue on the cancellation mechanism of ultraviolet (UV) divergences 
in a realistic gauge-Higgs unification model. 
We have considered $(D+1)$ dimensional $SU(3)$ gauge-Higgs unification model compactified 
on an orbifold $S^1/Z_2$ with a massive bulk fermion in a fundamental representation. 
The orbifolding is indispensable to obtain the SM Higgs $SU(2)_L$ doublet 
since Higgs originally behaves as an adjoint representation of the gauge group in the gauge-Higgs unification. 
In the case of $S^1/Z_2$, 
the bulk mass parameter of fermion must have odd $Z_2$ parity 
since the fermion bulk mass term connects fermions with different chiralities and opposite $Z_2$ parities. 
It is well known that the zero mode wave functions take an exponential profile along a compactified space coordinate 
and $D$-dimensional effective Yukawa couplings obtained by an overlap integral of zero mode wave functions 
are exponentially suppressed. 
In this way, we can freely obtain the light fermion masses, which are otherwise of ${\cal O}(M_W)$, 
by tuning the bulk mass parameter. 
One might worry if our argument for the finiteness in the above QED case  
still holds in the present orbifold model since the on-shell condition for the fermion is changed to 
$i \Gamma^M \langle D_M \rangle \Psi = M \varepsilon(y) \Psi$ 
($\varepsilon(y):$ the sign function of $y$, the extra space coordinate) and also the brane localized operator 
\begin{align}
i \bar{\psi}_L \Gamma^{\mu\nu} A_y \Gamma^y \psi_R F_{\mu\nu} 
\label{braneAMM}
\end{align}
seems to be allowed. 
However, these two worries are not necessary. 
As for the first one, we note that the fermion $\Psi$ in the operator (\ref{MMMD}) 
should be understood as the zero mode fermion. 
Though the operator (\ref{MMMD}) does not vanish even after imposing the on-shell condition, 
the remaining operator $M \varepsilon(y) \bar{\Psi} \Gamma^{MN} \Psi F_{MN}$ has no correspondence 
in the standard model (in the standard model $\bar{\Psi}_L \gamma^{MN} \Psi_R$ is not gauge invariant), 
and therefore vanishes automatically for the zero-mode fermion $\Psi$. 
As for the second one, note that the shift symmetry 
$A_y \to A_y + {\rm const}$ is operative as a remnant of higher dimensional gauge symmetry
even at the branes \cite{GIQ}. 
Therefore, the brane localized operator (\ref{braneAMM}) is forbidden by the shift symmetry. 
Furthermore, the UV finiteness is independent of how we compactify the extra space, 
because the information about the compactification is an infrared property of the theory. 
From these observations, we can expect the magnetic moment still to be finite 
even for the orbifold compactification and the presence of bulk mass term. 
We have shown \cite{ALM2} that the divergences due to the exchanges of nonzero KK gauge bosons and those due to the exchanges of its scalar partners exactly cancel, 
as the result of the Higgs-like mechanism for nonzero KK modes. 
As for zero modes, the divergences due to W, Z boson exchanges   
and those due to their scalar partner cancel since the ordinary Higgs mechanism works, 
but the cancellations of divergences due to photon and Higgs exchanges are incomplete 
since the corresponding partners are projected out by orbifold boundary conditions 
and the Higgs mechanism does not work.
It was emphasized that the contributions from photon and Higgs exchange become finite 
in five and six dimensional case, 
which should be contrasted with other higher dimensional models such as UED.  

In this paper, we focus on a five dimensional case of the model adopted in the previous paper \cite{ALM2}.  
The purposes are two folds. 
One is to show explicitly that Schwinger's result for the anomalous magnetic moment is reproduced, 
which could not be realized in \cite{ALM1} and was untouched in \cite{ALM2}. 
The other is to calculate finite nonzero KK mode contributions to $g-2$ and 
obtain a lower bound for the compactification scale by comparing with the experimental data.

This paper is organized as follows. 
In the next section, 
we introduce our model and discuss the mass eigenvalues and mode functions of fermions and gauge bosons. 
In section 3, we derive various interaction vertices and Feynman rules, 
which are needed in the calculation of the anomalous magnetic moment. 
In section 4, we calculate the muon anomalous magnetic moment, 
and show that Schwinger's result is reproduced and 
obtain a lower bound for the compactification scale from the finite nonzero KK mode contributions. 
Our conclusions are given in section 5.  
The detailed derivation of the vertex functions and amplitudes are summarized in Appendices A, B and C. 



\section{The Model}
We consider a five dimensional $SU(3)$ gauge-Higgs unification model 
compactified on an orbifold $S^1/Z_2$ with a radius $R$ of $S^1$. 
As a matter field, 
a massive bulk fermion in the third-rank totally symmetric tensor 
({\bf 10} dimensional) representation of $SU(3)$ gauge group is introduced, 
which has a SM lepton doublet and a singlet. 
The Lagrangian is given by
\begin{equation}
 \mathcal{L}
=
 -\frac{1}{2}\tr (F_{MN}F^{MN})+\bar{\Psi}({\bf 10})(i\not \!\!D -M\epsilon(y))\Psi({\bf 10})
\end{equation}
where the indices $M,N=0,1,2,3,5$, the five dimensional 
gamma matrices are $\Gamma^M=(\gamma^{\mu},i\gamma^{5})$
($\mu=0,1,2,3$),
\begin{align}
 F_{MN}
=&
 \partial_MA_N-\partial_NA_M -ig[A_M,A_N],
 \\
 \not \!\! D 
=&
 \Gamma^M (\partial_M -igA_M),
 \\
 \Psi({\bf 10})
=& \Psi({\bf 4}) \oplus \Psi({\bf 3}) \oplus \Psi({\bf 2}) \oplus \Psi({\bf 1}) \\
=&
 \left(\begin{array}{l}\Delta^{++} \\ \Delta ^+ \\ \Delta^0 \\\Delta^- \end{array}\right)
 +\left(\begin{array}{l}\Sigma^+\\\Sigma^0\\\Sigma^-\end{array}\right)
 +\left(\begin{array}{l} \Psi_1\\ \Psi_2\end{array}\right)
 +\Psi_3
\end{align}
where $g$ denotes a gauge coupling constant in five dimensional gauge theory.
$M$ is a bulk mass of the fermion. 
$\epsilon(y)$ is the sign function of an extra coordinate $y$ which is necessary 
to introduce a $Z_2$ odd bulk mass term. 
It is useful to decompose the fermion into representations of $SU(2) \times U(1)$ and 
the superscript on each fermionic field denotes its electric charge. 
$\left({\Psi_1\atop \Psi_2}\right),\Psi_3$ corresponds to 
Standard Model doublet and singlet, respectively.

The periodic boundary condition is imposed along $S^1$ for all fields and $Z_2$ parity
assignments are taken as
\begin{eqnarray}
A_{\mu}
&=&
 \left(
 \begin{array}{ccc}
  (+,+)&(+,+)&(-,-)\\
  (+,+)&(+,+)&(-,-)\\
  (-,-)&(-,-)&(+,+)
 \end{array}
 \right),
 ~~
 A_y
=
 \left(
 \begin{array}{ccc}
  (-,-)&(-,-)&(+,+)\\
  (-,-)&(-,-)&(+,+)\\
  (+,+)&(+,+)&(-,-)
 \end{array}
 \right), \label{BC1}
 \\
 \Psi({\bf 10})
&=&
(\Delta_L(+,+) + \Delta_R(-,-) ) \oplus
(\Sigma_L(-,-) + \Sigma_R(+,+) ) \nonumber \\
&&\oplus 
\left(
\begin{array}{c}
{\Psi_1}_L(+,+) + {\Psi_1}_R(-,-) \\
{\Psi_2}_L(+,+) + {\Psi_2}_R(-,-) \\ 
\end{array}
\right) \oplus
({\Psi_3}_L(-,-)+{\Psi_3}_R(+,+))
\label{BC2}
\end{eqnarray}
where (+,+) stands for $Z_2$ parity at fixed points $y=0, \pi R$.

\subsection{The mass eigenvalues and mode functions of gauge boson}

Let us first derive the mass eigenvalues and mode functions for gauge bosons $A_{\mu}$ 
and their scalar partners $A_y$.
They can be expanded in KK modes such that the boundary conditions (\ref{BC1}) and (\ref{BC2})are satisfied, 
\begin{align}
 A_{\mu}(x,y)
=&
 \sum_{n=1}^{\infty}
 \left(
 \begin{array}{ccc}
 (A^{3(n)}+\frac{1}{\sqrt{3}}A^{8(n)})C_n
  & (A^{1(n)}-iA^{2(n)})C_n
  & (A^{4(n)}-iA^{5(n)})S_n\\
 (A^{1(n)}+iA^{2(n)})C_n
  &(-A^{3(n)} + \frac{1}{\sqrt{3}}A^{8(n)})C_n
  & (A^{6(n)}-iA^{7(n)})S_n\\
 (A^{4(n)} + iA^{5(n)})S_n
  & (A^{6(n)} +iA^{7(n)})S_n
  &-\frac{2}{\sqrt{3}}A^{8(n)}C_n
 \end{array}
 \right)_{\mu}
 \notag
 \\ 
 &+
  \frac{1}{\sqrt{2\pi R}}
 \left(
 \begin{array}{ccc}
 A^{3(0)} + \frac{1}{\sqrt{3}}A^{8(0)}
  & A^{1(0)} -iA^{2(0)}
  & 0\\
 A^{1(0)} +i A^{2(0)} 
  &-A^{3(0)} + \frac{1}{\sqrt{3}}A^{8(0)}
  & 0\\
 0
  & 0
  &-\frac{2}{\sqrt{3}}A^{8(0)}  
 \end{array}
 \right)_{\mu}, 
 \\
 A_y(x,y)
=&
 \sum_{n=1}^{\infty}
 \left(
 \begin{array}{ccc}
 (A^{3(n)} + \frac{1}{\sqrt{3}}A^{8(n)})S_n
  & (A^{1(n)} -i A^{2(n)})S_n
  & (A^{4(n)} -i A^{5(n)})C_n \\
 (A^{1(n)} + iA^{2(n)})S_n
  &(-A^{3(n)} + \frac{1}{\sqrt{3}}A^{8(n)})S_n
  & (A^{6(n)} - iA^{7(n)})C_n\\
 (A^{4(n)} +i A^{5(n)})C_n
  & (A^{6(n)} +i A^{7(n)})C_n
  &-\frac{2}{\sqrt{3}}A^{8(n)} S_n
 \end{array}
 \right)_y
\notag\\
+&
 \frac{1}{\sqrt{2\pi R}}
 \left(
 \begin{array}{ccc}
  0 
  &0
  & A^{4(0)} -i A^{5(0)} \\
  0
  &0
  & A^{6(0)} -i A^{7(0)} \\
 A^{4(0)} +i A^{5(0)}
  & A^{6(0)} +i A^{7(0)}
  &0
 \end{array}
 \right)_y,
\end{align}
where $C_n=\frac{1}{\sqrt{\pi R}}\cos \left( \frac{n}{R}y \right), 
S_n =\frac{1}{\sqrt{\pi R}} \sin \left( \frac{n}{R}y \right)$.
Taking account of the Higgs VEV $\langle A_y\rangle$, quadratic terms 
relevant to gauge boson masses are diagonalized by
rotating gauge boson fields.
\begin{equation}
\begin{aligned}
 &{\mathcal L}_{\rm mass}\\
=
 &-\iy {\rm Tr}F_{\mu y}F^{\mu y}  - \frac{1}{2\xi}\iy
 [\partial^{\mu}A^a_{\mu} -\xi (\partial_y A_{y}^a-2m_W f^{6ab}A_{by})]^2
 \\
=&\sum_{n=1}^{\infty} 
 \frac{1}{2}A^{a(n)}_{\mu}M_{ab}^2 A^{b\mu(n)}
 -\frac{1}{2\xi}(\partial^{\mu}A_{a\mu}^{(n)})^2
 -\frac{\xi}{2}A_{y}^{a(n)}M'^2_{ab}A_{y}^{b(n)}
\\
=&
 \sum_{n=1}^{\infty}
 \Bigg[
  \frac{1}{2}M_n^2 (\gamma_{\mu}^{(n)}\gamma^{\mu(n)}+h_{\mu}^{(n)}h^{\mu (n)})
  +\frac{1}{2}(M_n-2m_W)^2\phi_{\mu}^{(n)}\phi^{\mu(n)}
  +\frac{1}{2}(M_n+2m_W)^2Z_{\mu}^{(n)}Z^{\mu(n)}
  \\
&
  +(M_n+m_W)^2W_{\mu}^{+(n)}W^{-\mu(n)}
  +(M_n-m_W)^2X_{\mu}^{+(n)}X^{-\mu(n)}
 \Bigg]
  +\frac{1}{2}(2m_W)^2Z_{\mu}Z^{\mu}+m_W^2W^+_{\mu}W^{-\mu}
\\
&
 -\xi \sum_{n=1}^{\infty}
 \Bigg[
  \frac{1}{2}M_n^2 (\gamma_y^{(n)}\gamma^{(n)}_y+h_y^{(n)}h^{(n)}_y)
  +\frac{1}{2}(M_n+2m_W)^2\phi_y^{(n)}\phi^{(n)}_y
  +\frac{1}{2}(M_n-2m_W)^2Z_y^{(n)}Z^{(n)}_y
  \\
&
  +(M_n-m_W)^2W_y^{+(n)}W^{-(n)}_y
  +(M_n+m_W)^2X_y^{+(n)}X^{-(n)}_y
 \Bigg]
  +\frac{1}{2}(2m_W)^2Z_yZ_y+m_W^2W^+_yW^-_y
\end{aligned}
\end{equation}
where the gauge-fixing term is introduced to eliminate the mixing terms 
between the gauge bosons and the gauge scalar bosons. 
The 't Hooft-Feynman gauge $\xi=1$ is adopted throughout this paper. 
 
The mass matrices $M,M',M_n,m_W$ are defined as:
\begin{align}
M=&
\left[
 \begin{array}{cccccccc}
 M_{11}^2 & 0 & 0 & 0 & M_{15}^2 & 0 & 0 & 0 \\
 0 & M_{22}^2 & 0 & M_{24}^2 & 0 & 0 & 0 & 0 \\
 0 & 0 & M_{33}^2 & 0  & 0 & 0 & M_{37}^2 & M_{38}^2 \\
 0 & M_{42}^2 & 0 & M_{44}^2 & 0 & 0 & 0 & 0 \\
 M_{51}^2 & 0 & 0 & 0 & M_{55}^2 & 0 & 0 & 0 \\
 0 & 0 & 0 & 0 & 0 & M_{66}^2 & 0 &  0 \\
 0 & 0 & M_{73}^2 & 0 & 0 & 0 & M_{77}^2 & M_{78}^2 \\
 0&0& M_{83}^2&0  & 0 & 0 & M_{87}^2 & M_{88}^2  
 \end{array}
\right],\\
M'=&M|_{m_W\to -m_W} 
\end{align}
where nonvanishing elements are
\begin{eqnarray}
&&M_{11}^2 = M_{22}^2 = M_{33}^2 = M_{44}^2 = M_{55}^2 = M_n^2+m_W^2, \quad M_{66}^2 = M_n^2, \nonumber \\
&&M_{77}^2 = M_n^2 + 4m_W^2, \quad M_{88}^2 = M_n^2 + 3m_W^2, \nonumber \\
&&M_{24}^2 = M_{42}^2 = M_{37}^2 = M_{73}^2 = -M_{15}^2 = -M_{51}^2 = 2m_W M_n, \nonumber \\
&&M_{38}^2 = M_{83}^2 =-\sqrt{3} m_W^2, \quad M_{78}^2 = M_{87}^2 = -2 \sqrt{3} m_W M_n, 
\end{eqnarray}
and $M_n=\frac{n}{R},m_W=2g\langle A_y^6\rangle = \frac{g}{\sqrt{2\pi R}}v=g_4 v$. 
$g_4$ is a four dimensional gauge coupling. 
The KK mass eigenstates $\gamma^{(n)},h^{(n)},\cdots$ 
and zero mode mass eigenstates $\gamma ,h ,\cdots$ are found as follows.
\begin{equation}
\begin{array}{ll}
\gamma^{(n)}
=\frac{1}{2}(\sqrt{3}A^{3(n)} + A^{8(n)}),
&h^{(n)}=A^{6(n)}
\\ 
Z^{(n)}=\frac{1}{\sqrt{2}}
 \left[\frac{A^{3(n)} - \sqrt{3}A^{8(n)}}{2} -A^{7(n)} \right],
&\phi^{(n)}=\frac{1}{\sqrt{2}}
 \left[\frac{A^{3(n)} - \sqrt{3}A^{8(n)}}{2} + A^{7(n)} \right],
\\
W^{\pm(n)}=\frac{1}{2}
 \left[A^{1(n)} + A^{5(n)} \mp i(A^{2(n)} - A^{4(n)}) \right],
&X^{\pm(n)}
=\frac{1}{2}
 \left[A^{2(n)} + A^{4(n)} \mp i (-A^{1(n)} + A^{5(n)}) \right].
\\
\gamma_{\mu}=\frac{1}{2}\left[\sqrt{3}A^3_{\mu} + A^8_{\mu} \right]
&h=A_{y}^{6(0)}
\\
W^{\pm}_{\mu}=\frac{1}{\sqrt{2}}(A^1_{\mu} \mp i A^2_{\mu}),
&X^{\pm}=\frac{1}{\sqrt{2}}\left[A^4_{y}\mp i A^5_{y} \right]
\\
Z_{\mu}=\frac{1}{2}(A^3_{\mu}-\sqrt{3}A^8_{\mu})
&\phi=A^7_{y}
\end{array}
\end{equation}
The zero mode gauge bosons $W^{\pm}_{\mu},Z_{\mu},\gamma_{\mu}$
correspond to $W$ boson, $Z$ boson and photon, respectively and zero mode scalar fields 
$X^{\pm},\phi,h$ correspond to charged NG boson, neutral NG boson, and
Higgs field in the Standard Model, respectively.

Some comments on this model are in order. 
First, the predicted Weinberg angle of this model is not realistic, $\sin^2 \theta_W = 3/4$. 
Possible way to cure the problem is to introduce an extra $U(1)$ or 
the brane localized gauge kinetic term \cite{SSS}. 
If we consider the case of additional $U(1)'$ case, 
we note that the mass eigenstates in the neutral sector 
are modified as
\begin{equation}
\begin{array}{ll}
\gamma^{(n)}
= \sin \theta_W A^{3(n)} + \cos \theta_W(\sin \varphi A^{8(n)} + \cos \varphi A^{'(n)}), \\
Z^{(n)}=\frac{1}{\sqrt{2}}
 \left[\cos \theta_W A^{3(n)} - \sin \theta_W (\sin \varphi A^{8(n)} + \cos \varphi A^{'(n)}) -A^{7(n)} \right], \\
\phi^{(n)}=\frac{1}{\sqrt{2}}
 \left[\cos \theta_W A^{3(n)} - \sin \theta_W (\sin \varphi A^{8(n)} + \cos \varphi A^{'(n)}) + A^{7(n)} \right],
\\
\gamma_{\mu}=\sin \theta_W A^{3} + \cos \theta_W(\sin \varphi A^{8} + \cos \varphi A^{'}), \\
Z_{\mu}=\cos \theta_W A^{3} - \sin \theta_W (\sin \varphi A^{8} + \cos \varphi A^{'}), \\
\phi=A_y^7 
\end{array}
\end{equation}
where $\theta_W$ is Weinberg angle of the Standard model and 
$\sin \varphi \equiv g_4'/\sqrt{3g_4^2 + (g_4')^2}, \cos \varphi \equiv \sqrt{3}g_4/\sqrt{3g_4^2 + (g_4')^2}$ 
in terms of gauge couplings of $SU(2)$ and extra $U(1)'$. 
The remaining linear combination of $A^{8(n)}$ and $A^{'(n)}$ 
\begin{equation}
A_X^{(n)} = \cos \varphi A^{8(n)} - \sin \varphi A^{'(n)}
\end{equation}
corresponds to the gauge boson of extra $U(1)_X$ symmetry 
unbroken after electroweak symmetry breaking. 

The effects of introducing extra $U(1)'$ appear in $A^{8(n)}, A^{'(n)}$ sector as
\begin{eqnarray}
&&A^{8(n)} = \cos \varphi A_X^{(n)} + \sin \varphi \left( \cos \theta_W \gamma^{(n)} -\sin \theta_W \frac{Z^{(n)}+\phi^{(n)}}{\sqrt{2}} \right), \\
&&A^{'(n)} = -\sin \varphi A_X^{(n)} + \cos \varphi \left( \cos \theta_W \gamma^{(n)} -\sin \theta_W \frac{Z^{(n)}+\phi^{(n)}}{\sqrt{2}} \right).
\end{eqnarray}
This shows that the gauge couplings of neutral sector are suppressed 
by the mixing angles $\sin \varphi, \cos \varphi$ 
compared to those in the $SU(3)$ model.  
Therefore, even if we take into account the correct Weinberg angle 
by introducing an extra $U(1)'$ gauge group, 
our order estimation is found not to be changed so much
since the contribution due to the exchange of the vector partner of Higgs, 
which is not modified by the additional $U(1)'$, dominates 
those due to the neutral gauge boson exchange, 
which are ${\cal O}(10^{-2})$ smaller as is shown in the table below eq. (4.7).

Second, the neutrino remains massless and we have no neutrino Yukawa coupling. 
A possible way out of this problem is to introduce a right-handed neutrino 
and to employ the seesaw mechanism to generate neutrino masses. 

\subsection{The mass eigenvalues and mode functions of fermion}

In this subsection, 
we derive fermion mass eigenvalues and mode functions. 
In our previous paper \cite{ALM2}, 
the mass eigenvalues and mode functions were obtained by solving equations of motion 
for fermion including mass term originated from $\langle A^6_y \rangle$.  
The advantages of this approach are that the amplitude of anomalous magnetic moment 
can be expressed in an exact form and it helps in analyzing the cancellation mechanism of divergences. 
However, the vertex functions are quite complicated and cannot be analytically solved, 
unless some approximation is applied or some extreme cases are considered. 
From this reason, we take a different approach in this paper. 
We first solve equations of motion for fermion without $\langle A^6_y \rangle$. 
Then, regarding the $\langle A^6_y \rangle$ originated mass term as a perturbation, 
we diagonalize the fermion mass matrix at a linear order of perturbation 
by making an approximation that the compactification scale is larger than the W-boson mass. 

As will be described in Appendix D, 
the contribution to the anomalous magnetic moment by $\Delta(\bf 4)$ vanishes 
and that by $\Sigma(\bf 3)$ is strongly suppressed 
compared with those by the SM doublet and singlet contributions. 
Therefore, we focus on the doublet and singlet parts in {\bf 10} dimensional representation 
and derive mass eigenvalues, 
the corresponding eigenfunctions and calculate their contributions to muon anomalous magnetic moment 
in the main text.

Expanding the doublet and singlet part assembled in a fermion denoted as $\Psi(x,y)$ 
in terms of parity even (odd) mode functions 
$f_L^{(n)}, f_L^{(0)},f_R^{(n)}, f_R^{(0)}(g^{(n)})$, 
\begin{equation}
 \Psi(x,y)
=
 \sum_{n=1}^{\infty}
 \left(
 \begin{aligned}
  \Psi^{(n)}_{1L}(x)f_L^{(n)}(y)
   +\Psi^{(n)}_{1R}(x)g^{(n)}(y)\\
  \Psi^{(n)}_{2L}(x)f_L^{(n)}(y)
   +\Psi^{(n)}_{2R}(x)g^{(n)}(y)\\
  \Psi^{(n)}_{3L}(x)g^{(n)}(y)
   +\Psi^{(n)}_{3R}(x)f_R^{(n)}(y)
 \end{aligned}
 \right)
+
 \left(
 \begin{aligned}
  \Psi^{(0)}_{1L}(x)f_L^{(0)}(y)\\
  \Psi^{(0)}_{2L}(x)f_L^{(0)}(y)\\
  \Psi^{(0)}_{3R}(x)f_R^{(0)}(y)
 \end{aligned}
 \right)
\end{equation}
where $n\geq 1$, 
the eigenvalue equations we should solve are given as
\begin{align}
&0=[\partial_5 +M\epsilon(y)]f_L^{(0)},\\
&0=[\partial_5 -M\epsilon(y)]f_R^{(0)},\\
&0=m_nf_R^{(n)}-[\partial_5 +M\epsilon(y)]f_L^{(n)},\\
&0=m_nf_L^{(n)}+[\partial_5 -M\epsilon(y)]f_R^{(n)},
\end{align}
where $m_n$ denotes the masses of $n$-th KK mode fermion mass.
The mode functions for zero modes $f_{L(R)}^{(0)}$ are obtained immediately.
\begin{equation}
 f^{(0)}_L=\sqrt{\frac{M}{1-\e ^{-2\pi RM}}}\e^{-M|y|}
 ~~,~~
 f^{(0)}_R=\sqrt{\frac{M}{\e^{2\pi RM}-1}}\e^{M|y|}.
\end{equation}
For nonzero KK modes, imposing the boundary conditions for parity even fields
\begin{equation}
\begin{aligned}
\partial_5 f_R^{(n)}(y)|_{y=0,\pi R} = M f_R^{(n)}(y)|_{y=0, \pi R}, 
\\
\partial_5 f_L^{(n)}(y)|_{y=0, \pi R} = -M f_L^{(n)}(y)|_{y=0, \pi R},
\end{aligned}
\end{equation}
and those for parity odd fields 
\begin{equation}
g^{(n)}(y)|_{y=0, \pi R}=0,
\end{equation}
we obtain the mode functions corresponding to $m_n=\sqrt{M^2+M_n^2}$.
\begin{align}
 f_L^{(n)}
=&
 \frac{M_n}{\sqrt{\pi R}m_n}
  \left[\c \left( \frac{n}{R}y \right) - \frac{MR}{n} \epsilon(y)\s \left( \frac{n}{R}y \right) \right],
 \\
 f_R^{(n)}
=&
 \frac{M_n}{\sqrt{\pi R}m_n}
  \left[ \c \left( \frac{n}{R}y \right) + \frac{MR}{n}\epsilon(y)\s \left( \frac{n}{R}y\right) \right],
 \\
 g^{(n)}
=&
 \frac{1}{\sqrt{\pi R}}\s \left( \frac{n}{R}y \right).
\end{align}
We can see that fermion mass term without electroweak symmetry breaking is diagonalized.
\begin{equation}
\begin{aligned}
\iy
  \bar{\Psi}(i\Gamma^5\partial_5 -M\epsilon(y))\Psi
=&
 \sum_{n=1}^{\infty} m_n[
 \bar{\Psi}_1^{(n)} \Psi_1^{(n)} + \bar{\Psi}_2^{(n)} \Psi_2^{(n)} - \bar{\Psi}_3^{(n)} \Psi_3^{(n)} ]
\\
\to &
 -\sum_{n=1}^{\infty}m_n
 [\bar{\Psi}_1^{(n)} \Psi_1^{(n)} + \bar{\Psi}_2^{(n)} \Psi_2^{(n)} + \bar{\Psi}_3^{(n)} \Psi_3^{(n)} ]
\end{aligned}
\end{equation}
In the second line, the chiral rotation 
$\Psi_{1,2}\to \e ^{i\pi \gamma_{5}/2}\Psi_{1,2}$
is performed.

From now on, we focus on fermion mass matrix taking account of Higgs VEV 
$\langle A_y\rangle$.
Since Standard Model Higgs field $h$ is regarded as $A_y^6$ in our model, 
it produces mixing terms between $\Psi_2$ and $\Psi_3$ after electroweak symmetry breaking. 
(Note that $A_y^6$ couples to only $\Psi_2,\Psi_3$.)
The fermion mass matrix is diagonalized as;
\begin{align}
 &\bar{\Psi}_L M \Psi_R +{\rm h.c.}
 \notag
\\
=&
 (\bar{\Psi}_{2L}^{(0)},\bar{\Psi}_{2L}^{(1)},\bar{\Psi}_{3L}^{(1)},
 \bar{\Psi}_{2L}^{(2)},\bar{\Psi}_{3L}^{(2)},\ldots)
 \left(
 \begin{array}{cccccc}
  m_{\mu}& 0& -\tilde{m}_1 & 0 & -\tilde{m}_2& \\
  -\hat{m}_1&m_1 &-\tilde{m}_{W1} & 0 & -\tilde{m}_{12}&\\
  0 & -m_W & m_1 &0  &0&\\
  -\hat{m}_2&0&-\tilde{m}_{21}& m_2& -\tilde{m}_{W2}&\\
  0&0&0&-m_W&m_2&\\
  &&&&&\ddots
 \end{array}
 \right)
 \left(
 \begin{array}{c}
 \Psi_{3R}^{(0)}\\ \Psi_{2R}^{(1)}\\ \Psi_{3R}^{(1)}
\\ \Psi_{2R}^{(2)}\\ \Psi_{3R}^{(2)}
 \\ \vdots
 \end{array}
 \right)+({\rm h.c.})
 \notag\\
=&
 (\bar{\mu}_L,\bar{\psi}_{2L}^{(1)}, \bar{\psi}_{3L}^{(1)}\cdots)
 \left(
 \begin{array}{cccc}
 m_\mu & & & \\
 & m_1^- & & \\
 & & m_1^+ & \\
 & & & \ddots \\
 \end{array}
 \right)
 \left(
 \begin{array}{c}
 \mu_R\\ \psi_{2R}^{(1)}\\ \psi_{3R}^{(1)}
 \\ \vdots
 \end{array}
 \right)
 +\mathcal{O}(R^2m_W^2)
\end{align}
where 
\begin{equation}
\begin{array}{ll}
m_{\mu}=\frac{2\pi RM}{\sqrt{(1-\e^{-2\pi RM})(\e^{2\pi RM}-1)}}m_W,
&\tilde{m}_{Wn}=\left(1-\frac{2M^2}{m_n^2}\right)m_W
\\
\hat{m}_n=4\sqrt{\frac{\pi RM}{1-\e^{-2\pi RM}}}
          \frac{1-(-1)^n\e^{-\pi RM}}{\pi Rm_n^3}M_nM
&\tilde{m}_n=(-1)^n\hat{m}_n
\\
\tilde{m}_{nl}=\frac{4nl(1-(-1)^{n+l})}{\pi Rm_nm_l(n^2-l^2)}(1-\delta_{nl})m_W M, 
&(m_n^\pm)^2 = m_n^2 \pm 2m_W \frac{M_n^2}{m_n}. 
\end{array}
\end{equation}
The mass eigenstates of fermion
$\mu_R,\mu_L,\psi^{(n)}_2,\psi^{(n)}_3$ 
are obtained as:
\begin{align}
 \mu_L
=& 
 \Psi^{(0)}_{2L}+\sum_{n=1}^{\infty} \frac{\hat{m}_n}{m_n}\Psi^{(n)}_{3L}
  ,~~~
  \mu_R=\Psi^{(0)}_{3R}
 +\sum_{n=1}^{\infty}(-1)^n\frac{\hat{m}_n}{m_n}\Psi^{(n)}_{3R},
\\
 \psi_{3L}^{(n)}
=&
 \frac{1}{\sqrt{2}}\left[
 \Psi_{2L}^{(n)}+\Psi_{3L}^{(n)}
 +\frac{M^2}{2m_n^3}m_W(\Psi_{2L}^{(n)}-\Psi_{3L}^{(n)})
 -\frac{\hat{m}_n}{m_n}\Psi_{2L}^{(0)}
 +\sum_{l\neq n}^{\infty} \frac{\tilde{m}_{nl}}{m_n^2-m_l^2}
  (m_l\Psi^{(l)}_{3L}-m_n\Psi_{2L}^{(l)})
 \right],
 \\
 \psi_{2L}^{(n)}
=&
 \frac{1}{\sqrt{2}}\left[
 \Psi_{2L}^{(n)}-\Psi_{3L}^{(n)}
 -\frac{M^2}{2m_n^3}m_W(\Psi_{2L}^{(n)}-\Psi_{3L}^{(n)})
 +\frac{\hat{m}_n}{m_n}\Psi_{2L}^{(0)}
 +\sum_{l\neq n}^{\infty} \frac{\tilde{m}_{nl}}{m_n^2-m_l^2}
  (m_l\Psi^{(l)}_{3L}+m_n\Psi_{2L}^{(l)})
 \right],
 \\
 \psi_{3R}^{(n)}
=&
 \frac{1}{\sqrt{2}}\left[
 \Psi_{2R}^{(n)}+\Psi_{3R}^{(n)}
 -\frac{M^2}{2m_n^3}m_W(\Psi_{2L}^{(n)}-\Psi_{3L}^{(n)})
 -(-1)^n\frac{\hat{m}_n}{m_n}\Psi_{2L}^{(0)}
 +\sum_{l\neq n}^{\infty} \frac{\tilde{m}_{nl}}{m_n^2-m_l^2}
  (m_n\Psi^{(l)}_{3R}-m_l\Psi_{2R}^{(l)})
 \right],
 \\
 \psi_{2R}^{(n)}
=&
 \frac{1}{\sqrt{2}}\left[
 \Psi_{2R}^{(n)}-\Psi_{3R}^{(n)}
 -\frac{M^2}{2m_n^3}m_W(\Psi_{2R}^{(n)}-\Psi_{3R}^{(n)})
 +\frac{\hat{m}_n}{m_n}\Psi_{2R}^{(0)}
 +\sum_{l\neq n}^{\infty} \frac{\tilde{m}_{nl}}{m_n^2-m_l^2}
  (m_l\Psi^{(l)}_{3R}+m_n\Psi_{2R}^{(l)})
 \right],
 \\
\nu_L=& \Psi_{1L}^{(0)}, \quad \psi_1^{(n)}=\Psi_1^{(n)}. 
\end{align}
Note that $\Psi^{(n)}_1,\Psi^{(0)}_1$ remain unchanged 
since Higgs $A_y^{(0)}$ only couples to $\Psi_2$ and $\Psi_3$.
The zero mode fields $\mu$ and $\nu$ correspond to muon and muon neutrino in the Standard Model. 
Putting these mode functions and integrating over a fifth coordinate, 
we obtain a quadratic part of 4D effective Lagrangian of fermion
\begin{equation}
\begin{aligned}
 \mathcal{L}
=&
 \sum_{n=1}^{\infty}
 \left[
  \bar{\psi}_1^{(n)}(i \partial\!\!\!/ -m_n)\psi_1^{(n)}
  +\bar{\psi}_2^{(n)}(i \partial\!\!\!/ -m_n^-)\psi_2^{(n)}
  +\bar{\psi}_3^{(n)}(i\partial\!\!\!/ -m_n^+)\psi_3^{(n)}
 \right]\\
 &+\bar{\mu}(i \partial\!\!\!/ -m_{\mu})\mu +\bar{\nu}_L \partial\!\!\!/ \nu_L. 
\end{aligned}
\end{equation}

\section{Interaction vertices and its Feynman rules}

In this section, 
we derive interaction vertices 
to calculate muon anomalous magnetic moment.
The necessary interaction vertices to calculate one loop contribution
to $\gamma_{\mu}\bar{\mu}\gamma^{\mu}\mu$ are three point vertices 
where the one of the external fermions is zero mode,
 or the external photon is zero mode. 
\begin{figure}[h]
\begin{equation*}
\begin{array}{c}
\begin{array}{c}
\includegraphics{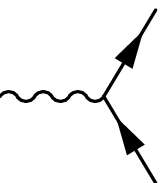}
\begin{picture}(0,0)(1,1)
\put (5,0){$\mu,\psi^{(n)}_{2,3}$}
\put (5,50){$\mu,\psi^{(n)}_{2,3}$}
\put (-62,25){$\gamma_{\mu}$}
\end{picture}
\end{array}
=-e\gamma_{\mu}
,~~~~
\begin{array}{c}
\includegraphics{fig1.eps}
\begin{picture}(0,0)(1,1)
\put (5,0){$\mu$}
\put (5,50){$\mu,\psi_{2,3}^{(n)}$}
\put (-64,25){$\gamma_{\mu}^{(m)}$}
\end{picture}
\end{array}
=-e\gamma_{\mu}V^{(m)}_{\gamma}(\mu,\psi^{(n)}_{2,3})
,~~~~\begin{array}{c}
\includegraphics{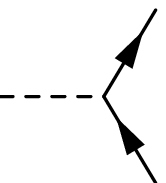}
\begin{picture}(0,0)(1,1)
\put (5,0){$\mu$}
\put (5,50){$\psi_{2,3}^{(n)}$}
\put (-64,25){$\gamma_y^{(m)}$}
\end{picture}
\end{array}
=-eU^{(m)}_{\gamma}(\psi^{(n)}_{2,3}).
\end{array}
\end{equation*}
\label{photonint}
\caption{Feynman rules whose external gauge boson is photon,
photon KK mode, photon NG boson.}
\end{figure}

The interaction vertices of photon 
$\gamma_{\mu}$ and photon scalar partner $\gamma_y$ are derived as follows (see Figure 1).
\begin{align}
&
ig \iy\bar{\Psi}
 (A_{\mu}\gamma^{\mu}+A_y)\Psi \notag \\
&\supset 
 -ie \left[ \gamma_{\mu}^{(0)}\bar{\mu} \gamma_{\mu}\mu
 + \gamma_{\mu}^{(0)}\bar{\psi}_{2,3}^{(n)} \gamma^{\mu}\psi_{2,3}^{(n)}
 + \gamma_{\mu}^{(m)}
  \bar{\mu} V^{(m)}_{\gamma}\gamma^{\mu}\mu
 + \gamma_{\mu}^{(m)}\bar{\psi}_{2,3}^{(n)} V^{(m)}_{\gamma}\gamma^{\mu}\mu
 + \gamma_y^{(m)}\bar{\psi}_{2,3}^{(n)} U_\gamma(\psi_{2,3})\mu 
 \right]
 \label{3.1}
\end{align}
where 
the muon couplings of the photon are extracted. 
As for the $A_y$ coupling, a chiral rotation 
$\psi \to e^{-i \gamma^y \pi/4} \psi$ is performed to get rid of $\gamma^y$ factor. 
The electric charge is given by the weak gauge coupling $g_4$ and 
the Weinberg angle $\sin^2 \theta_W=3/4$ in the present model 
through  $e= g_4 \sin \theta_W=\frac{\sqrt{3}}{2}g_4$. 
The vertex functions in (\ref{3.1}) are explicitly given by
\begin{align}
V^{(m)}_{\gamma}(\mu)
 =&
 \sqrt{2\pi R}
 I_{cLL}^{0m0}
 [L+(-1)^mR],
 \\
 V^{(m)}_{\gamma}(\psi^{(n)}_{2,3})
=&
 \sum_{l=1}^{\infty} \sqrt{\pi R}I_c^{lmn}
 \frac{\hat{m}_l}{m_l}[\mp L+(-1)^lR]
  - 
\frac{\hat{m}_n}{m_n}
 \sqrt{\pi R}I_{cLL}^{0m0}
 [\mp L +(-1)^{n+m}R], 
 \notag\\
&
 +
 I_{cRR}^{0mn}
 \Bigg[
 (-1)^{n+m}\left(1\mp \frac{M^2}{M_n^3}m_W\pm 
 \sum_{l\ne n}^{\infty}\frac{m_l\tilde{m}_{nl}}{m_n^2-m_l^2}\right)L
 \notag\\
 &
 +\left(\mp 1 -\frac{M^2}{2m_n^3}m_W
 +\sum_{l\ne n}^{\infty}\frac{m_l\tilde{m}_{nl}}{m_n^2-m_l^2}\right)R
 \Bigg]
 \\
U^{(m)}_{\gamma}(\psi_{2,3}^{(n)})
=&
 -i\sqrt{\pi R} \sum_{l=1}^\infty
 \left[
 -\frac{\tilde{m}_n}{m_n}
 I_{sL}^{lmn}L
 \pm \frac{\hat{m}_l}{m_l}I_{sR}^{lmn}R
 \right]
 \notag\\
&
 -i\sqrt{\pi R}I_{sR}^{0mn}
 \left(
 \mp 1 -\frac{M^2}{2m_n^3}m_W +\sum_{l\neq n}^{\infty}
 \frac{m_n\tilde{m}_{nl}}{m_n^2-m_l^2} \right)
 (-R\mp (-1)^{n+m}L).
\end{align}
The sign $\pm$ means that if the internal nonzero KK mode fermion is $\psi_2(\psi_3)$,
we take a plus (minus) sign. 
Also $L$ and $R$ denote chiral projection operators defined as
$L=\frac{1+\gamma_{5}}{2}, R=\frac{1-\gamma_{5}}{2}$. 
The integrals of mode functions $I_{cLL}^{0m0}$ etc are summarized in Appendix A.3. 
It is straightforward to derive other interaction vertices and Feynman rules necessary 
for the calculation of muon $g-2$ and their results are also summarized in Appendix A. 

\section{Calculation of muon anomalous magnetic moment}

Now, we are ready to calculate muon $g-2$. 
Photon coupling at the tree level is modified due to the quantum correction as
\begin{equation}
-e\bar{\mu}(\gamma^{\mu}+\Gamma^{\mu})\mu.
\end{equation}
Among a few terms in $\bar{\mu}\Gamma^{\mu}\mu$,
we are interested in the term proportional to $(p^{\mu}+p'^{\mu})$
with a form factor $F_2(0)$.
\begin{equation}
\bar{\mu}(p')\Gamma^{\mu}\mu(p)\to
\bar{\mu}(p')
\left[
-\frac{1}{2m_{\mu}}(p^{\mu}+p'^{\mu})F_2(0)
\right]\mu(p)
\end{equation}
where $p,p'$ denotes incoming, outgoing momentum of external muon, respectively.
It is the form factor $F_2(0)$ that gives the muon anomalous magnetic moment:
$a=\frac{g-2}{2}=F_2(0)$.

\subsection{Schwinger's result}

In a paper by the present author \cite{ALM1}, 
we have shown that the anomalous magnetic moment is finite in a toy model of gauge-Higgs QED. 
Although this result itself is quite striking, 
we could not reproduce Schwinger's famous result of anomalous magnetic moment 
from zero mode diagram. 
This is because Higgs exchange contribution is comparable to photon one due to the fact 
that Yukawa coupling is given by the gauge coupling in the toy model of gauge-Higgs unification.

In the present model, 
the undesirable features can be avoided by introducing a bulk mass term $M\epsilon(y)$.
We show that the famous Schwinger's result concerning electron anomalous magnetic moment
is reproduced in the present model.
\vspace{1mm}
\begin{align}
\begin{array}{c}
\includegraphics[scale=0.8]{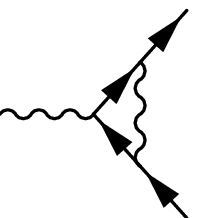}
\begin{picture}(0,0)
\put(0,-4){$\mu(p)$}
\put(0,48){$\mu(p')$}
\put(-60,22){$\gamma_{\mu}$}
\put(-7,22){$\gamma_{\mu}$}
\put(-28,36){$\mu$}
\put(-28,4.8){$\mu$}
\end{picture}
\end{array}
&= -e^3 \int \frac{d^Dk}{(2 \pi)^D i} 
\frac{\gamma_\nu (p'\!\!\!\!/ + k\!\!\!/ + m_\mu) \gamma^\mu (p\!\!\!/ + k\!\!\!/ + m_\mu) \gamma^\nu}
{[(p'+k)^2-m_\mu^2][(p+k)^2-m_\mu^2]k^2} 
\supset \frac{e^3}{16\pi^2} \frac{1}{m_\mu} (p^\mu+p'^\mu)
\nonumber \\
&\equiv
 eF_2^{\gamma_{\mu}}(0)\frac{1}{2m_{\mu}}(p^{\mu}+p'^{\mu})
 \Rightarrow F^{\gamma_{\mu}}_2(0)=\frac{e^2}{8\pi^2}. 
 \\
 \notag \\
\begin{array}{c}
\includegraphics[scale=0.8]{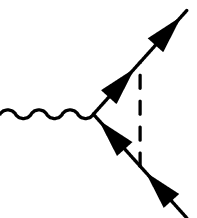}
\begin{picture}(0,0)
\put(0,-4){$\mu$}
\put(0,48){$\mu$}
\put(-60,22){$\gamma_{\mu}$}
\put(-7,22){$h_y$}
\put(-28,36){$\mu$}
\put(-28,4.8){$\mu$}
\end{picture}
\end{array}
&= -3e^3 \left( \frac{m_\mu}{m_W} \right)^2 
\int \frac{d^D k}{(2\pi)^Di} 
\frac{(p\!\!\!/' + k\!\!\!/ + m_\mu) \gamma^\mu (p\!\!\!/ + k\!\!\!/ + m_\mu)}
{[(p'+k)^2 - m_\mu^2][(p+k)^2 - m_\mu^2][k^2-m_H^2]} \nonumber \\
&\supset \frac{3e^3 m_\mu}{2\pi^2 m_H^2} \left( \frac{m_\mu}{m_W} \right)^2 \ln \left( \frac{m_\mu}{m_H} \right)^2 
(p^\mu+p'^\mu) \nonumber \\
&\equiv
 eF_2^{h_y}(0)\frac{1}{2m_{\mu}}(p^{\mu}+p'^{\mu})
 \Rightarrow F^{h_y}_2(0)
\simeq
 3\left(
 \frac{e}{\pi}\frac{m_{\mu}}{m_H}\frac{m_{\mu}}{m_W}
 \right)^2
 \ln \left(\frac{m_{\mu}}{m_H}\right)^2. 
\end{align}
The contribution from Higgs exchange diagram is strongly suppressed 
compared with the photon exchange diagram.
Thus, Schwinger's result is obtained. 

\subsection{The contribution from nonzero KK mode}

In this subsection, 
we calculate one loop contributions from 
nonzero KK modes to the magnetic moment $a(\rm KK)$ numerically.
In our calculations, 
an approximation $R m_W  \ll 1$ 
and numerical values 
\begin{equation}
\begin{aligned}
&\frac{m_{\mu}}{m_W}
=\frac{2\pi RM}{\sqrt{(1-\e^{-2\pi RM})(\e^{2\pi RM}-1)}}\Rightarrow
MR\sim 3.05,
\\
&\alpha=\frac{e^2}{4\pi}\sim\frac{1}{137}
\end{aligned}
\end{equation}
are used. 
In the text, only the results are shown, 
but the details of calculation are summarized in Appendices. 


The contribution from neutral current sector $a({\rm N.C.})$ contains 
those from photon exchange $a(\gamma_{\mu})$,
the exchange of photon scalar partner $a(\gamma_y),\cdots$,{\it i.e.},
\begin{equation}
	 a({\rm N.C.})
	=
	 a(\gamma_{\mu})+a(\gamma_y)
	+a(h_{\mu})+a(h_y)
	+a(Z_{\mu})+a(Z_y)
	+a(\phi_{\mu})+a(\phi_y)
	\end{equation}
from the diagrams of type
	\begin{equation}
	\begin{aligned}
	 &
	 \begin{array}{c}
	 \includegraphics[scale=0.8]{fig6.eps}
	 \end{array}
	 \supset
	 \frac{e}{2m_{\mu}}F_2(0)(p^{\mu}+p'^{\mu}). 
	\end{aligned}
	\end{equation}
Numerical results of each contribution (in unit of $(m_W R)^2$) are summarized in a table below. 
	\begin{center}
	\begin{tabular}{c|c|c|c}
	$a(\gamma_{\mu})$ & $a(\gamma_y)$ & $a(h_{\mu})$ &$a(h_y)$ 
	\\
	\hline
	$-1.01\times 10^{-8}$ & $3.21\times 10^{-7}$ & $1.21\times 10^{-5}$& $-2.25\times 10^{-8}$ 
	\\
	\hline\hline
	$a(Z_{\mu})$ & $a(Z_y)$ & $a(\phi_{\mu})$ & $a(\phi_y)$
	\\
	\hline
	$8.63 \times 10^{-8}$ &  $3.22 \times 10^{-7}$ & $8.63\times 10^{-8}$ & $3.22 \times 10^{-7}$ 
	\end{tabular}
	\end{center}
Thus, we have the contribution from neutral current sector, $a(\rm N.C.)$.
\begin{equation}
a({\rm N.C.})=1.32\times 10^{-5} (Rm_W)^2. 
\label{neutral}
\end{equation}


Next, we turn to charged current sector (C.C.).
The contribution from charged current sector $a(\rm C.C.)$ is
calculated as 
\begin{equation}
a({\rm C.C.})=-5.02\times 10^{-7} (Rm_W)^2. 
\label{charged}
\end{equation}
Thus, combining the results (\ref{neutral}) and (\ref{charged}) leads to 
the contribution of nonzero KK modes $a(\rm KK)$ 
\begin{equation}
\label{result}
a({\rm KK})
=
1.27\times 10^{-5}(Rm_W)^2.
\end{equation}


Finally, we focus on the comparison of our result (\ref{result})
with experimental result from E821 at Brookhaven National Laboratory \cite{BNL}.
Our result $a(\rm KK)$ must be less than the difference 
between the experimental data $a(\rm Exp)$ and Standard Model prediction $a(\rm SM)$:
\begin{equation}
a({\rm KK})< a({\rm Exp})-a({\rm SM})=(2.90\pm 0.90)\times 10^{-9},
\end{equation}
which gives a lower bound for the compactification scale  
\begin{equation}
\frac{1}{R} > (58 \sim 80)m_W=4.7 \sim 6.4~{\rm TeV}. 
\end{equation}
This lower bound is quite natural to realize a viable Higgs boson mass in the flat space gauge-Higgs unification. 
Finite Higgs mass is radiatively generated and is of the order $m_H^2 \sim \frac{\alpha}{4\pi}(1/R)^2$. 
For the Higgs mass to be of order 100 GeV, 
the compactification scale is required to be of order ${\cal O}({\rm TeV})$. 


\section{Conclusion}

In this paper, we have investigated the muon anomalous magnetic moment (muon $g-2$) 
in a five dimensional $SU(3)$ gauge-Higgs unification model compactified on an orbifold $S^1/Z_2$. 
As a matter field, a fermion in the ten dimensional representation 
with $Z_2$ parity odd bulk mass term is introduced. 
In \cite{ALM1}, Schwinger's result could not be obtained 
since Higgs exchange diagram was comparable to photon one 
due to the fact that muon Yukawa coupling is given by the gauge coupling. 
But in this paper, muon Yukawa coupling is obtained by choosing the bulk mass appropriately. 
As a result, 
the contribution from Higgs exchange to $g-2$ is strongly suppressed 
compared to the photon exchange, {\it i.e.},
ordinary QED correction and famous Schwinger's result $a=\frac{e^2}{8\pi}$
is reproduced.

We have also calculated the contribution from nonzero KK modes and 
obtained a lower bound for the compactification scale 
by comparing with the experimental data at Brookhaven National Laboratory (BNL).
Especially, there exists a difference between the experimental data and
Standard Model prediction with a significance of 3.2 $\sigma$ standard deviations. 
Assuming that nonzero KK mode contributions are responsible for the deviation, 
we obtain a lower bound for the compactification scale around 5 $\sim$ 6 TeV.

\subsection*{Acknowledgments}

The work of the authors was supported 
in part by the Grant-in-Aid for Scientific Research 
of the Ministry of Education, Science and Culture, No.18204024 and No. 20025005.  

\appendix

\section{Various Feynman rules}
\label{variousfeynmanrules}

We summarize various Feynman rules to calculate 
muon anomalous magnetic moment. 
Substituting mode expansions of $\Psi,A_{\mu},A_y$ 
into the gauge (scalar) coupling $\bar{\Psi}A_M\gamma^{M}\psi$ 
and integrating over the fifth coordinate,
$4D$ effective gauge (scalar) interaction vertices are obtained.

\subsection{Neutral current}

\def\beq {\begin{equation}}
\def\eeq {\end{equation}}
\def\bea {\begin{eqnarray}}
\def\eea{\end{eqnarray}}
\def\half{\frac{1}{2}}

In this subsection, 
we summarize the couplings of neutral gauge boson 
and their scalar partners 
$\gamma_{\mu,y},Z_{\mu,y},h_{\mu,y},\phi_{\mu,y}$.
We describe the vertex function as follows.
\begin{equation}
\begin{array}{c}
\includegraphics[scale=0.8]{fig1.eps}
\end{array}
\begin{picture}(0,0)(0.1,0.1)
\put(-3,-15){$\mu$}
\put(-3,25){$\mu,\psi_{2,3}^{(n)}$}
\put(-68,0){$A_{\mu}^{(m)}$}
\end{picture}
=-e\gamma_{\mu}V^{(m)}_{A}(\mu,\psi_{2,3}^{(n)})
,~~~~~~
\begin{array}{c}
\includegraphics[scale=0.8]{fig2.eps}
\end{array}
\begin{picture}(0,0)(0.1,0.1)
\put(-3,-15){$\mu$}
\put(-3,25){$\mu,\psi^{(n)}_{2,3}$}
\put(-70,0){$A_y^{(m)}$}
\end{picture}
=-e
U^{(m)}_A(\mu,\psi_{2,3}^{(n)}).
\end{equation}
The vertex fuctions of photon and photon scalar partner have already been described in section 3. 

As for Higgs ($h_{\mu,y}$), the vertex functions $V,U$ are given as 
obtained.
\begin{align}
 V_h^{(m)}(\psi_{2,3}^{(n)})
=&
 \sqrt{\pi R}I_{sR}^{0mn}
 \Bigg[
  (-1)^{n+m}\left(\mp 1 -\frac{M^2}{2m_n^3}m_W 
   +\sum_{l\neq n}^{\infty}\frac{m_n\tilde{m}_{nl}}{m_n^2-m_l^2}\right)L \nonumber 
\\&
  +\left(1\pm \frac{M^2}{2m_n^3}m_W 
         \pm \sum_{l\neq n}^{\infty}\frac{m_n\tilde{m}_{nl}}{m_n^2-m_l^2}\right)R
 \Bigg] \nonumber \\
&
+ \sum_{l=1}^\infty \frac{\hat{m}_l}{m_l}\sqrt{3\pi R}
 [I_{sL}^{lmn}L \mp (-1)^{n+m}I_{sR}^{lmn}R], 
\\
 V_h^{(m)}(\mu)
=&
 \sqrt{6\pi R} \sum_{n=1}^\infty I_{sR}^{0mn}\frac{\tilde{m}_n}{m_n}
 \left[
  (-1)^mL+R
 \right], \\
 U_h^{(0)}(\mu)
=&
 i\sqrt{3}I_{LR}^{00}, 
 \\
 U_h^{(0)}(\psi_{2,3}^{(n)})
=&
 i\frac{\sqrt{6}}{2}I_{LR}^{0n}
 \Bigg[
  \left(\pm 1 -\frac{M^2}{2m_n^3}m_W 
   -\sum_{l\neq n}^\infty\frac{m_l\tilde{m}_{nl}}{m_n^2-m_l^2}\right)L
 \nonumber \\
&
 +(-1)^n
 \left(
  1\mp \frac{M^2}{2m_n^3}m_W\pm \sum_{l\ne n}^\infty\frac{m_l\tilde{m}_{nl}}{m_n^2-m_l^2}
 \right)R
 \Bigg]
 \nonumber \\
&
 +i\frac{\sqrt{6}}{2}I_{LR}^{00} \frac{\hat{m}_n}{m_n}[(-1)^n L \pm R]
 -i\frac{\sqrt{6}}{2} \frac{\hat{m}_n}{m_n}[L\pm (-1)^n R], 
 \\
 U^{(m)}_h (\psi_{2,3}^{(n)})
=&
 i\sqrt{2\pi R}I_{cLR}^{nm0}
 \Bigg[
  (-1)^{n+m}\left(\pm 1 -\frac{M^2}{2m_n^3}m_W 
  - \sum_{l \ne n}^\infty \frac{m_l\tilde{m}_{nl}}{m_n^2-m_l^2}\right)L
  \nonumber \\
&
 +\left(1\mp \frac{M^2}{2m_n^3}m_W \pm \sum_{l\ne n}^\infty \frac{m_l\tilde{m}_{nl}}{m_n^2-m_l^2}\right)R
 \Bigg] \nonumber \\
& -i\sqrt{\pi R} \sum_{l=1}^\infty I_c^{lmn} \frac{\hat{m}_l}{m_l}
 [L\pm (-1)^lR]. 
\end{align}
As for $Z_{\mu,y}$ and $\phi_{\mu,y}$ bosons,
we note that there is a simple relation. 
Focusing on the gauge interactions concerning $Z, \phi$, we have
\begin{equation}
 \bar{\Psi} A_M \Psi|_{Z,\phi,\Psi_{2,3}}
\supset 
 \sqrt{2}(\bar{\Psi}_2,\bar{\Psi}_3)
 \left(\begin{array}{cc}
  \phi_M + Z_M & -i(\phi_M - Z_M) \\
  i(\phi_M - Z_M) & \phi_M + Z_M
 \end{array}\right)
 \left(\begin{array}{c}
  \Psi_2\\ \Psi_3
 \end{array}\right)
\end{equation}
which tells us that the vertex functions of $Z_M^{(m)}$ corresponds to 
the complex conjugate of those of $\phi_M^{(m)}$, 
$V^{(m)}_{\phi}(\psi^{(n)}_{2,3})=[V^{(m)}_{Z}(\psi_{2,3}^{(n)})]^{\ast}$, 
for instance. 
The vertex functions of $Z_M^{(m)}$ and $\phi_M^{(m)}$ are thus given by
\begin{align}
 V_Z^{(m)}(\psi_{2,3}^{(n)})
=&
 \frac{\sqrt{6}}{2}\sqrt{\pi R}
 \Bigg[
  (-1)^{n+m}\left(1\mp \frac{M^2}{2m_n^3}m_W 
  \pm \sum_{l\ne n}^\infty \frac{m_n\tilde{m}_{nl}}{m_n^2-m_l^2}\right)L
 \nonumber \\
&+\left(\pm 1 -\frac{M^2}{2m_n^3}m_W - \sum_{l\ne n}^\infty \frac{m_n\tilde{m}_{nl}}{m_n^2-m_l^2}\right)R
 \Bigg]
 \nonumber \\
&-\frac{\sqrt{6}}{2}\sqrt{\pi R}I_{sR}^{0mn}
\Bigg[
 (-1)^{n+m}\left(\mp 1 - \frac{M^2}{2m_n^3}m_W + \sum_{l\ne n}^\infty \frac{m_n\tilde{m}_{nl}}{m_n^2-m_l^2}\right)L
 \nonumber \\
&+\left(
 -1\pm \frac{M^2}{2m_n^3}m_W \mp \sum_{l\ne n}^\infty \frac{m_n\tilde{m}_{nl}}{m_n^2-m_l^2}
 \right)R
\Bigg]
\nonumber \\
&-\frac{\sqrt{6}}{2}\sqrt{\pi R}I_{cLL}^{0m0}\frac{\hat{m}_n}{m_n}
 [\mp L -(-1)^{n+m}R]
-\frac{\sqrt{6}}{2}\sqrt{\pi R} \sum_{l=1}^\infty I_c^{lmn}
 \frac{\hat{m}_l}{m_l}[\mp L -(-1)^l R]
 \nonumber \\
&
-i \frac{\sqrt{6}}{2}\sqrt{\pi R}
 \sum_{l=1}^\infty [I_{sL}^{lmn}L \pm (-1)^l I_{sR}^{lmn}R], \\
 V_Z^{(0)}(\psi_{2,3}^{(n)})
=&
 -\sqrt{6}\frac{\hat{m}_n}{m_n}[\mp L -(-1)^nR], \\
 V_Z^{(0)}(\mu)
=&
 \sqrt{3}(L-R),
\end{align}
\begin{align}
 U_{\phi}^{(m)}(\psi_{2,3}^{(n)})
=&
 i\frac{\sqrt{6}}{2}\sqrt{\pi R}I_{sL}^{0mn}
 \Bigg[
 \left(1\pm \frac{M^2}{2m_n^3}m_W \pm \sum_{l\ne n}^\infty \frac{m_n\tilde{m}_{nl}}{m_n^2-m_l^2}\right)L
 \nonumber \\
&
 +(-1)^{n+m}
 \left(\mp 1 -\frac{M^2}{2m_n^3}m_W + \sum_{l\ne n}^\infty \frac{m_n\tilde{m}_{nl}}{m_n^2-m_l^2}\right)R
 \Bigg]
\nonumber \\
&-\frac{\sqrt{6}}{2} \sqrt{\pi R}\sum_{l=1}^\infty \frac{\hat{m}_l}{m_l} 
\left[
i(\pm I_{sR}^{lmn}L -(-1)^l I_{sL}^{lmn}R)
+ I_c^{lmn}
(L\mp (-1)^l R) 
\right] 
\nonumber \\
&-\frac{\sqrt{6}}{2}\sqrt{\pi R}(-1)^{n+m}I_{cLR}^{nm0}
\Bigg[
 \left(\pm 1 -\frac{M^2}{2m_n^3}m_W + \sum_{l\ne n}^\infty \frac{m_l\tilde{m}_{nl}}{m_n^2-m_l^2}\right)L
 \nonumber \\
&-(-1)^{n+m}\left(1\mp \frac{M^2}{2m_n^3}m_W \pm \sum_{l \ne n}^\infty \frac{m_l\tilde{m}_{nl}}{m_n^2-m_l^2}\right)R
\Bigg], \\
 U_{\phi}^{(m)}(\mu)
=&
 i\sqrt{3\pi R} \sum_{n=1}^\infty I_{sL}^{0mn}
 \frac{\hat{m}_n}{m_n}(-1)^n (1-(-1)^m)(L-R), \\
 U_{\phi}^{(0)}(\psi^{(n)}_{2,3})
=&
 \frac{\sqrt{6}}{2}\frac{\hat{m}_n}{m_n}[L\mp (-1)^n R]
 +\frac{\sqrt{6}}{2}I_{LR}^{0n} \frac{\hat{m}_n}{m_n}
  [(-1)^nL \mp R]
 \nonumber \\
 &
 -\frac{\sqrt{6}}{2}I_{LR}^{00}
 \Bigg[
 \left(\pm 1 -\frac{M^2}{2m_n^3}m_W - \sum_{l\neq n}^{\infty} \frac{m_l\tilde{m}_{nl}}{m_n^2-m_l^2}
 \right)L 
 \nonumber \\
 &
 - (-1)^n\left(1\mp \frac{M^2}{2m_n^3}m_W \pm \sum_{l\neq n}^{\infty} \frac{m_l\tilde{m}_{nl}}{m_n^2-m_l^2}
 \right)R
 \Bigg], \\
 U_{\phi}^{(0)}(\mu)
=&
 -\sqrt{3}I_{LR}^{00}. 
\end{align}

\subsection{Charged current}

Concerning about charged gauge bosons such as $W_{\mu,y}^{\pm},X_{\mu,y}^{\pm}$,
we can write down general form of vertex functions as
\vspace*{5mm}
\begin{equation}
\begin{array}{c}
\includegraphics[scale=0.8]{fig1.eps}
\end{array}
\begin{picture}(0,0)(0.1,0.1)
\put(-3,-15){$\mu$}
\put(-3,25){$\nu,\psi_1^{(n)}$}
\put(-68,0){$A_{\mu}^{+(m)}$}
\end{picture}
=-e\gamma_{\mu}V^{(m)}_A(\nu,\psi_{1}^{(n)})
,~~~~~~~~~~
\begin{array}{c}
\includegraphics[scale=0.8]{fig2.eps}
\end{array}
\begin{picture}(0,0)(0.1,0.1)
\put(-3,-15){$\mu$}
\put(-3,25){$\nu,\psi^{(n)}_1$}
\put(-70,0){$A_y^{+(m)}$}
\end{picture}
=-e
U^{(m)}_A(\nu,\psi_{1}^{(n)})
\end{equation}
where the various vertex functions are given by
\begin{align}
 V_W^{(m)}(\psi_1^{(n)})
=&
 \sqrt{6\pi R}\left[
 I_{cLL}^{0mn}L-I_{sR}^{0mn}R
 \right]
 +\sqrt{6\pi R} \sum_{l=1}^\infty \frac{\hat{m}_l}{m_l}
 [I_{sL}^{lmn}L +(-1)^l I_c^{lmn}R]
 \nonumber \\
&
+\sqrt{6\pi R}\left[
 I_{cLL}^{0m0}+\frac{\hat{m}_n}{m_n}I_{sR}^{0mn}
 \right]L,
 \\
 V_W^{(0)}(\psi^{(n)}_1)
=&
 \sqrt{6}\frac{\tilde{m}_n}{m_n}R, 
 \\
 V_W^{(0)}(\nu)
=&
 \sqrt{6}L,
 \\
 V_{X}^{(m)}(\psi_1^{(n)})
=&
  i\sqrt{6\pi R}\left[
 I_{cLL}^{0mn}L+I_{sR}^{0mn}R
 \right] 
 +i\sqrt{6\pi R} \sum_{l=1}^\infty \frac{\hat{m}_l}{m_l}
 [-I_{sL}^{lmn}L +(-1)^l I_c^{lmn}R]
 \nonumber \\
&
+i\sqrt{6\pi R}\left[
 I_{cLL}^{0m0}-\frac{\hat{m}_n}{m_n}I_{sR}^{0mn}
 \right]L, 
\end{align}
\begin{align}
 U_X^{(m)}(\psi_1^{(n)})
=&
 -\sqrt{6\pi R}\left[
  I_{sL}^{0mn}L -I_{cLR}^{nm0}R
 \right]
+\sqrt{6\pi R} \sum_{l=1}^\infty \frac{\hat{m}_l}{m_l}
 \left[
   I_c^{lmn}L + I_{sL}^{lmn}R
 \right],
 \\
 U_X^{(m)}(\nu)
=&
 \sqrt{6\pi R}\frac{\tilde{m}_n}{m_n}I_{sL}^{0mn}R,
 \\
 U_X^{(0)}(\nu)
=&
 \sqrt{6}I_{LR}^{00}R,
 \\
 U_W^{(m)}(\psi_1^{(n)})
=&
  i\sqrt{6\pi R}\left[
  I_{sL}^{0mn}L +I_{cLR}^{nm0}R
 \right]
+i\sqrt{6\pi R} \sum_{l=1}^\infty \frac{\hat{m}_l}{m_l}
 \left[
  -I_c^{lmn}L + I_{sL}^{lmn}R
 \right],
\end{align}

\subsection{Integrals of mode functions}

Here is a summary of integrals of mode functions appearing in various vertex functions.
\begin{align}
I_{LR}^{00} \equiv&\iy f_L^{(0)}f_R^{(0)}, \quad I_{LR}^{0n} \equiv \iy f_L^{(0)}f_R^{(n)},\quad  I_{LR}^{ln}\equiv\iy f_L^{(l)}f_R^{(n)}\\
I_{cLL}^{0m0}\equiv&\iy f_L^{(0)} C_m f_L^{(0)}, \quad 
I_{cRR(LL)}^{0mn}\equiv \iy f_{R(L)}^{(0)}C_mf_{R(L)}^{(n)}, \\
I_{cLR}^{nm0} \equiv &\iy f_L^{(n)}C_mf_R^{(0)}, \quad I_{sR(L)}^{0mn}\equiv \iy f_{R(L)}^{(0)} S_mg^{(n)}, \\
I_{sL(R)}^{lmn}\equiv &\iy g^{(l)}S_m f_{L(R)}^{(n)}, \quad I_c^{lmn}\equiv \iy g^{(l)}C_mg^{(n)},\\
I_{cLR}^{lmn}\equiv& \iy f_L^{(l)}C_mf_R^{(n)}.
\end{align}

\subsection{Gauge boson self coupling}

\begin{figure}[h]
\begin{align*}
&\begin{array}{c}
\includegraphics[scale=0.8]{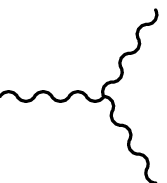}
\begin{picture}(0,0)(1,1)
\put(5,-5){$W_{\rho}^+,W_{\rho}^{+(n)},X_{\nu}^{+(n)}$}
\put(5,33){$W_{\nu}^{-},W_{\nu}^{-(n)},X_{\nu}^{-(n)}$}
\put(-5,26){\rotatebox[origin=c]{235}{$\to$}}
\put(-5,12){\rotatebox[origin=c]{125}{$\to$}}
\put(-30,25){$\to$}
\put(-52,20){$\gamma_{\mu}$}
\end{picture}
\end{array}
~~~~~~=
-3e[(q-p_+)_{\rho}\eta_{\mu\nu}+(p_+-p_-)_{\mu}\eta_{\nu\rho}+(p_- -q)_{\nu}\eta_{\mu\rho}]
~~~~~~(n\geq 1),
\\
&
\\
&
\begin{array}{c}
\includegraphics[scale=0.8]{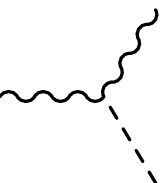}
\begin{picture}(0,0)(1,1)
\put(5,0){$W_{y}^{+(n)}$}
\put(5,33){$X_{\nu}^{-(n)}$}
\put(-5,26){\rotatebox[origin=c]{235}{$\to$}}
\put(-5,12){\rotatebox[origin=c]{125}{$\to$}}
\put(-30,25){$\to$}
\put(-52,20){$\gamma_{\mu}$}
\end{picture}
\end{array}
~~~=
-3e(M_n-m_W)\eta_{\mu\nu}~(n\geq 1),~~~~
\begin{array}{c}
\includegraphics[scale=0.8]{fig4.eps}
\begin{picture}(0,0)(1,1)
\put(5,0){$X_{y}^{+(n)}$}
\put(5,33){$W_{\nu}^{-(n)}$}
\put(-5,26){\rotatebox[origin=c]{235}{$\to$}}
\put(-5,12){\rotatebox[origin=c]{125}{$\to$}}
\put(-30,25){$\to$}
\put(-52,20){$\gamma_{\mu}$}
\end{picture}
\end{array}
~~~=3e(M_n+m_W)\eta_{\mu\nu}~(n\geq 0),
\\
&
\\
&\begin{array}{c}
\includegraphics[scale=0.8]{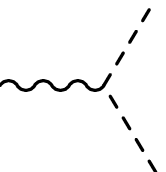}
\begin{picture}(0,0)(1,1)
\put(5,-5){$X_y^-,X_y^{-(n)},W_y^{-(n)}$}
\put(5,33){$X_y^{+},X_y^{+(n)},W_y^{+(n)}$}
\put(-5,26){\rotatebox[origin=c]{235}{$\to$}}
\put(-5,12){\rotatebox[origin=c]{125}{$\to$}}
\put(-30,25){$\to$}
\put(-52,20){$\gamma_{\mu}$}
\end{picture}
\end{array}
~~~=3e(p_+-p_-)_{\mu}~~~~(n\geq 1).
\end{align*}
\caption{Three point gauge vertices relevant for $g-2$.}
\label{selfinteraction}
\end{figure}

We also need self interactions of gauge bosons. 
Three point self interactions of gauge bosons shown in Fig. \ref{selfinteraction} are relevant for the calculation of $g-2$. 
$p_{+(-)}$ and $q$ denote momenta of charged gauge boson 
$W^{+(-)}_{\mu,y},X^{+(-)}_{\mu,y}$ and photon $\gamma_{\mu}$
and all momenta are taken such as to flow into the vertex.

\section{Numerical calculation of the magnetic moment}

\def\tr {{\rm Tr}}
\def\e {{\rm e}}
\def\c {\cos}
\def\s {\sin}
\def\d {{\rm d}}
\def\mot {\!\!\not \!}
\def\iy {\int_{-\pi R}^{\pi R}\!\!\!\!{\rm d}y}
\def\E{\frac{\pi RM}{{\rm e}^{2\pi RM}-1}}     
\def\D{\mathcal{D}}                            
\def\nt{\notag}                                
\def\C{\mathcal{C}}                             
\def\dX{\int_0^1\!\!\!{\rm d}X}
%
We are ready to calculate finite value of 
anomalous magnetic moment in this model.
To carry out this calculation, we use approximation that 
the ratio between muon mass and W-boson mass is 
much smaller than 1, {\it i.e.} $\frac{m_{\mu}}{m_W}\ll 1$.

\subsection{Neutral current sector}

The contributions to $g-2$ from neutral current sector are those 
due to the exchanges of photon, Higgs, 
the scalar partner of photon, the partner of Higgs, Z-boson and the scalar patner of Z-boson. 
The final results of the approximated amplitude and our numerical results are summarized below. 

\subsubsection{photon exchange}

\begin{align}
  a(\gamma_{\mu1})
=&
 \frac{2m_{\mu}}{e}\begin{array}{c}
  \includegraphics{fig6.eps}
  \begin{picture}(0,0)(1,1)
   \put(-8,30){$\gamma_{\mu}^{(n)}$}
   \put(-35,40){$\mu$}
   \put(-35,15){$\mu$}
  \end{picture}
 \end{array}
\sim
  \frac{e^2m_{\mu}^2}{2\pi^2}
  (2MR)^4R^2
    \sum_{n=1}^{\infty}
    \left( \frac{1}{(2MR)^2 +n^2}\right)^2
    \left( -\frac{2}{3} +(-1)^n \right)\frac{1}{n^2}. 
\end{align}

\begin{align}
 a(\gamma_{\mu2})
=&
 \frac{2m_{\mu}}{e}
 \begin{array}{c}
  \includegraphics{fig6.eps}
  \begin{picture}(0,0)(1,1)
   \put(-8,30){$\gamma_{\mu}^{(m)}$}
   \put(-50,40){$\psi_{2,3}^{(n)}$}
   \put(-50,15){$\psi_{2,3}^{(n)}$}
  \end{picture}
 \end{array}
 \nonumber \\
\sim&
 \frac{e^2(MR)^3}{\pi^3} R^2 m_W m_{\mu} 
\left[ 
 S_{\gamma_{\mu}}^{1}
 +
 \pi MR e^{-\pi RM}
S_{\gamma_{\mu}}^{2}
 -
2S_{\gamma_{\mu}}^{3}
 -
2S_{\gamma_{\mu}}^{4}
 +32
(MR)^2 S_{\gamma_{\mu}}^{5}
\right]. 
\end{align}
The mode sums are defined in the following.
\begin{align}
  S_{\gamma_{\mu}}^{1}
 =&
  \sum_{n,m=1}^\infty \dX
  \left[
  \frac{1}{R^2m_{n+m}^2}-\frac{1}{R^2m_{n-m}^2}
  \right]^2
  \frac{m^2n^2}{R^4m_n^4}
  \frac{(-1)^{n+m}(1-X)X}
       {\D}, \notag \\
  S_{\gamma_{\mu}}^{2}
 =&
  \sum_{n,m=1}^\infty \dX
  \frac{m^2}{R^2m_n^2}
  \left[
   \frac{1}{R^2m^2_{n+m}}-\frac{1}{R^2m_{n-m}^2}
  \right]^2
  \frac{X(1-X)(X-2)}{\D}, \notag \\
  S_{\gamma_{\mu}}^{3}
 =&
  \sum_{n,m=1}^\infty 
  \dX
  \left[
   \frac{1}{R^2m_{n+m}^2}-\frac{1}{R^2m_{n-m}^2}
  \right]^2
  \frac{m^2(-1)^{n+m}(1-X)X^2}{\D^2}
  \frac{n^2}{R^2m_n^2}, \notag \\
  S_{\gamma_{\mu}}^{4}
  =&
  2\sum_{n,m=1}^\infty \dX(-1)^{n+m}
  \left[
   \frac{1}{R^2m_{n+m}^2}-\frac{1}{R^2m_{n-m}^2}
  \right]
\notag \\
  &\hspace{5mm}\times
  \left[
   \frac{m(n+m)}{R^4m_{n+m}^4}+2\frac{m(n-m)}{R^4 m_{n-m}^4}
  \right]\frac{X(1-X)}{\D}, 
  \notag \\
  S_{\gamma_{\mu}}^{5}
 =&
  \sum_{n,m=1}^\infty
  \dX
  \frac{1}{(2MR)^2 +m^2}
  \frac{nm}{R^4m_n^4}
  \left[
   \frac{1}{R^2m_{n+m}^2}-\frac{1}{R^2m_{n-m}^2}
  \right]
  \frac{X(1-X)(-1)^{n+m}}{\D}
\end{align}
where the denominator $\D=[(MR)^2 +n^2-m^2]X +m^2$.

\subsubsection{Higgs exchange}

\begin{align}
 a(h_{y1})
=&
 \frac{2m_{\mu}}{e}
 \begin{array}{c}
  \includegraphics{fig7.eps}
 \begin{picture}(0,0)(1,1)
   \put(-8,30){$h_y$}
   \put(-50,40){$\psi_{2,3}^{(n)}$}
   \put(-50,15){$\psi_{2,3}^{(n)}$}
  \end{picture}
 \end{array}
 \nt\\
\sim& 
 -\frac{2e^2(MR)^3}{\pi^3}R^2m_Wm_{\mu} \times \nonumber \\
& \left[
  9 \sum_{n=1}^{\infty}\frac{(-1)^nn^4}{[(MR)^2 +n^2]^5}
+
14\pi MR e^{-\pi RM}
  \sum_{n=1}^{\infty}\frac{n^2}{[(MR)^2 + n^2]^4}
-
6 \sum_{n=1}^{\infty}\frac{(-1)^nn^2}{[(MR)^2 + n^2]^4}
\right],
\end{align}

\begin{align}
 a(h_{y2})
=&
 \frac{2m_{\mu}}{e}
 \begin{array}{c}
  \includegraphics{fig7.eps}
  \begin{picture}(0,0)(1,1)
   \put(-8,30){$h^{(m)}_y$}
   \put(-50,40){$\psi_{2,3}^{(n)}$}
   \put(-50,15){$\psi_{2,3}^{(n)}$}
  \end{picture}
 \end{array}
 \nt\\
\sim& 
 \frac{3e^2}{2\pi^3}(MR)^3R^2m_Wm_{\mu} 
 \left[ 
 S_{h_y}^{1} 
 - 
2\pi MR e^{-\pi RM}S_{h_y}^{2}
 +
(MR)^2 S_{h_y}^{3}
 -
4S_{h_y}^{4}
 -
2S_{h_y}^{5}
\right] 
\end{align}
where the mode sums are 
\begin{align}
  S_{h_y}^{1}
 =&
  \sum_{n,m=1}^{\infty}
 \dX X^2
 \frac{(-1)^{n+m}n^2}{R^4m_n^4}
 \left[
  \frac{2n+m}{R^2m_{n+m}^2}+\frac{2n-m}{R^2m_{n-m}^2}
 \right]^2
 \frac{1}{\D}, \notag \\
  S_{h_y}^{2}
 =&
  \sum_{n,m=1}^{\infty}\dX X^2
 \frac{1}{R^2m_n^2}
 \left[
   \frac{2n+m}{R^2m_{n+m}^2}+\frac{2n-m}{R^2m_{n-m}^2}
 \right]
 \frac{1-X}{\D}, \notag \\
  S_{h_y}^{3}
 =&
  \sum_{n,m=1}^{\infty}
 \dX X^2 \left[
  \frac{2n+m}{R^2m_{n+m}^2}+\frac{2n-m}{R^2m_{n-m}^2}
 \right]^2
 \frac{1}{\D}
 \frac{(-1)^{n+m}}{R^4m_n^4}, \notag \\
  S_{h_y}^{4}
 =&
  \sum_{n,m,l=1}^{\infty}
 (\delta_{n+m,l}+\delta_{m+l,n}-\delta_{n+l,m} )
 \dX X^2
\left[
  \frac{2n+m}{R^2m_{n+m}^2}+\frac{2n-m}{R^2m_{n-m}^2}
 \right]
 \frac{1}{n} \frac{m_n}{\D}
 \frac{l(-1)^l}{R^4m_l^4}, \notag \\
  \notag \\
  S_{h_y}^{5}
 =&
  \sum_{n,m=1}^{\infty}
  \dX X^2
  \frac{(-1)^{n+m}n^2}{R^2m_n^2}
 \left[
   \frac{2n+m}{R^2m_{n+m}^2}+\frac{2n-m}{R^2m_{n-m}^2}
  \right]^2
 \frac{1}{\D^2}. 
\end{align}

\subsubsection{The exchange of photon's partner}

\begin{align}
 a(\mathcal{N}_{\gamma_y})
=&
 \frac{2m_{\mu}}{e}
\begin{array}{c}
  \includegraphics{fig7.eps}
 \begin{picture}(0,0)(1,1)
   \put(-8,30){$\gamma^{(m)}_y$}
   \put(-50,40){$\psi_{2,3}^{(n)}$}
   \put(-50,15){$\psi_{2,3}^{(n)}$}
  \end{picture}
 \end{array}
 \nt\\
\sim&
 \frac{e^2(MR)^3}{4\pi^3}R^2 m_{\mu} \times \nonumber \\
& \left[
m_W S_{\gamma_y}^{1}
 -
2m_W S_{\gamma_y}^{2}
 -
(MR)^2 m_W S_{\gamma_y}^{3}
 +
m_{\mu} S_{\gamma_y}^{4}
 -
2m_W \left(
   S_{\gamma_y}^{5}
  +S_{\gamma_y}^{6} 
 \right)
 +
\frac{4MR}{\pi}
S_{\gamma_y}^{7}
\right] 
\end{align}
where
\begin{align}
 S_{\gamma_y}^{1}
=&
 \sum_{n,m=1}^\infty
 \int_0^1\!\!\!dXX^2
 \left[
  \frac{1}{R^2m_{n+m}^2}-\frac{1}{R^2m_{n-m}^2}
 \right]^2
 \frac{n^2}{R^2m_n^2}
 \frac{(-1)^{n+m}}{D}, \notag \\
 S_{\gamma_y}^{2}
=&
 \sum_{n,m=1}^\infty
 \int_0^1\!\!\!dXX^2
 \left[
  \frac{1}{R^2m_{n+m}^2}-\frac{1}{R^2m_{n-m}^2}
 \right]^2
 \frac{(-1)^{n+m}Xn^2}{D^2}, \notag \\
 S_{\gamma_y}^{3}
=&
 \sum_{n,m=1}^\infty
 \int_0^1\!\!\!dXX^2
 \left[
  \frac{1}{R^2m_{n+m}^2}-\frac{1}{R^2m_{n-m}^2}
 \right]^2
 \frac{(-1)^{n+m}}{R^2m_n^2}
 \frac{1}{D}, \notag \\
 S_{\gamma_y}^{4}
=&
 \sum_{n,m=1}^\infty
 \int_0^1\!\!\!dXX^2
 \left[
  \frac{1}{R^2m_{n+m}^2}-\frac{1}{R^2m_{n-m}^2}
 \right]^2
 \frac{X-1}{D}, \notag \\
 S_{\gamma_y}^{5}
=&
 \sum_{n,m=1}^\infty
 \int_0^1\!\!\!dXX^2
 (-1)^{n+m}
 \left[
  \frac{1}{R^2m_{n+m}^2}-\frac{1}{R^2m_{n-m}^2}
 \right]
 \left[
  2\frac{m-n}{R^4m_{n+m}^4}+\frac{m+n}{R^4m_{n-m}^4}
 \right] \frac{n}{D}, 
 \notag \\
 S_{\gamma_y}^{6}
=&
 \sum_{n,m=1}^\infty
 \int_0^1\!\!\!dXX^2
 (-1)^{n+m}
 \left[
  \frac{1}{R^2m_{n+m}^2}-\frac{1}{R^2m_{n-m}^2}
 \right]
 \left[
  2\frac{(n-m)^2}{R^5m_{n-m}^5}
  -\frac{(n+m)^2}{R^5m_{n+m}^5}
 \right] \frac{Rm_n}{D}, 
 \notag \\
 S_{\gamma_y}^{7}
=&
 \sum_{n,m,l=1}^\infty
 \int_0^1\!\!\!dXX^2
 \frac{l^2mn}{R^4m_l^4}(-1)^{n+m}
 (1-(-1)^{l+m+n})
 \left[
  \frac{1}{R^2m_{n+m}^2}-\frac{1}{R^2m_{n-m}^2}
 \right]\frac{Rm_n}{D} 
\notag \\
&\hspace{5mm}\times
 \left[
  \frac{1}{[(l+m)^2-n^2][(l-m)^2-n^2]} +\frac{1}{[(n+m)^2-l^2][(n-m)^2-l^2]}
 \right]. 
\end{align}

\subsubsection{The exchange of Higgs' partner}

\begin{align}
 a(h_{\mu})
=&
 \frac{2m_{\mu}}{e}
\begin{array}{c}
  \includegraphics{fig6.eps}
 \begin{picture}(0,0)(1,1)
   \put(-8,30){$h^{(m)}_{\mu}$}
   \put(-50,40){$\psi_{2,3}^{(n)}$}
   \put(-50,15){$\psi_{2,3}^{(n)}$}
  \end{picture}
 \end{array}
 \nt\\ 
\sim&
 \frac{3e^2}{2\pi^3}(MR)^3 R^2 m_{\mu}
 \left[ m_\mu S_{h_{\mu}}^{1}
 +
m_WS_{h_{\mu}}^{2}
 +
\frac{8}{\pi}m_W S_{h_{\mu}}^{3}
 +\frac{64}{\pi^2} m_W MR S^{4}_{h_{\mu}} 
 \right]
\end{align}
where
\begin{align}
  S_{h_{\mu}}^{1}
 &=
  \sum_{m,n=1}^\infty \int_0^1dX
 \frac{(1-X)(X-2)}{D}
 \left[
  \frac{1}{(Rm_{n+m})^2}- \frac{1}{(Rm_{n-m})^2}
 \right]^2, 
  \nt \\
  S_{h_{\mu}}^{2}\displaystyle
 &=
  \sum_{m,n=1}^\infty \int_0^1dX
  \frac{(1-X)(MR)^3}{X(Rm_n)^2+(1-X)m^2}
 \left[
  \frac{1}{(Rm_{n+m})^2}- \frac{1}{(Rm_{n-m})^2}
 \right]^2, 
 \nt \\
  S_{h_{\mu}}^{3}
&=
  \sum_{l,m,n=1}^{\infty}\int_0^1dX 
  \frac{(1-X)(MR)^3}{X(Rm_n)^2+(1-X)m^2}
 \left[
  \frac{1}{(Rm_{n+m})^2}- \frac{1}{(Rm_{n-m})^2}
 \right]
 \frac{n}{Rm_n}
 \nt  \\
&\hspace{5mm}~~~~~~~~~
\times 
 \left(
 \frac{2(m-n)}{(Rm_{m-n})^4}+\frac{m+n}{(Rm_{n+m})^4}
 \right)
 (-1)^{n+m}, 
 \nt \\
  S_{h_{\mu}}^{4}
&= \sum_{l,m,n=1}^\infty 
 \frac{(1-X)(MR)^4}{X(Rm_{n+m})^2 +(1-X)m^2}
 \left[
  \frac{1}{(Rm_{n+m})^2} - \frac{1}{(Rm_{n-m})^2}
 \right]
 \frac{1}{Rm_n}
 \notag \notag \\
&\hspace{5mm}~~~~~~~~~\times \displaystyle
 \frac{l^2mn((-1)^{n+m}-(-1)^l)}{[(l+m)^2-n^2][(l-m)^2 -n^2]}
 \frac{1}{(Rm_l)^4}. 
\end{align}

\subsubsection{Z boson exchange }

\begin{align}
 a(Z_{\mu1})
=&
 \frac{2m_{\mu}}{e}
\begin{array}{c}
  \includegraphics{fig6.eps}
 \begin{picture}(0,0)(1,1)
   \put(-8,30){$Z^{(m)}_{\mu}$}
   \put(-40,40){$\mu$}
   \put(-40,15){$\mu$}
  \end{picture}
 \end{array}
\sim
 \frac{12e^2}{\pi^2}(MR)^4
 \sum_{m=1}^{\infty}\frac{(-1)^m+\frac{2}{3}}{[(2MR)^2+m^2]^2m^2}
 (m_{\mu}R)^2,
 \\
 a(Z_{\mu2})
=&
 \frac{2m_{\mu}}{e}
 \begin{array}{c}
  \includegraphics{fig6.eps}
 \begin{picture}(0,0)(1,1)
   \put(-8,30){$Z^{(m)}_{\mu}$}
   \put(-50,40){$\psi_{2,3}^{(n)}$}
   \put(-50,15){$\psi_{2,3}^{(n)}$}
  \end{picture}
 \end{array}
 \notag \\
\sim &
 \frac{3e^2Rm_{\mu}(MR)^2}{4\pi^3}
 \Bigg[
  -(MR) Rm_{\mu}S^1_{Z_{\mu}}
  +2MR Rm_{\mu}S^2_{Z_{\mu}}
  -64(MR)^3 Rm_W S^3_{Z_{\mu}}
  \notag\\
&
  -8MR Rm_W S^4_{Z_{\mu}}
  -16 Rm_WS^5_{Z_{\mu}}
  + 2(MR)^3Rm_WS^6_{Z_{\mu}} 
  -\frac{4 R m_W}{(MR)^2} S^{7}_{Z_{\mu}} 
 \Bigg]
\end{align}
where
\begin{align}
  S^1_{Z_{\mu}}
 =& \sum_{n=1}^\infty 
  \dX \frac{X(1-X)(X-2)}{D}
  \left( \frac{n^2}{R^2m_n^2} +\frac{n}{M^2R^2}\right)
  \frac{1}{R^4m_n^4}, 
  \nt \\
  S^2_{Z_{\mu}}
 =& \sum_{m,n=1}^\infty 
  \dX \frac{X(1-X)}{D}
  \left(\frac{n^2}{R^2m_n^2}+\frac{n^2}{M^2R^2}\right)
  \frac{n^2(-1)^{n+m}}{R^6m_n^6}, 
  \nt \\ 
  S^3_{Z_{\mu}}
 = & \sum_{m,n=1}^\infty 
  \dX \frac{X(1-X)}{D}
  \frac{n^2}{R^6m_n^6}
  \frac{(-1)^n}{(2MR)^2+m^2}, 
  \nt \\ 
  S^4_{Z_{\mu}}
 =& \sum_{m,n=1}^\infty 
  \dX \frac{X(1-X)}{D}
  \frac{n(-1)^n}{R^2m_n^2}
  \left[2\frac{m-n}{R^4m_{n-m}^4}+\frac{n+m}{R^4m_{n+m}^4}\right], 
  \nt \\ 
  S^5_{Z_{\mu}}
 =&
  \frac{1}{2} \sum_{m,n=1}^\infty 
  \dX \frac{X(1-X)}{D}
  \frac{(-1)^nn^3}{R^2m_n^2}\frac{1}{R^4m_{n+m}^4}
  \nt \\
 &
  -\frac{4MR}{\pi} \sum_{l,m,n=1}^\infty \int dX \frac{X(1-X)}{D}
  \frac{(-1)^nn^2}{R^2m_n^2}\frac{lmn}{R^4m_l^4}
  \frac{1-(-1)^{l+m+n}}{[(l+m)^2-n^2][(l-m)^2-n^2]}, 
  \nt \\
  S^6_{Z_{\mu}}
 =& \sum_{m,n=1}^\infty 
  \dX \frac{X(1-X)}{D}
  \frac{n^2(-1)^{n+m}}{R^6m_n^6}, 
  \nt \\
  S^7_{Z_{\mu}}
 =& \sum_{m,n=1}^\infty
  \dX \frac{X(1-X)(-1)^{n+m}}{D^2}
  \left(\frac{n^2}{R^2m_n^2}+\frac{n^2}{M^2R^2}\right)
  \frac{n^2}{R^4m_n^4}. 
\end{align}

\subsubsection{$\phi_y$ exchange}

\begin{align}
 a(\phi_{y1})
 \notag
=&
\frac{2m_\mu}{e}
 \begin{array}{c}
  \includegraphics{fig7.eps}
 \begin{picture}(0,0)(1,1)
   \put(-8,30){$\phi^{(m)}_y$}
   \put(-40,40){$\mu$}
   \put(-40,15){$\mu$}
  \end{picture}
 \end{array}
 \nt\\
 \sim&
 -e^2\frac{6R^5m_{\mu}m_WM^3}{\pi^3}
 \sum_{n=1}^\infty \Bigg[
  -\frac{4}{3}\pi RM \e^{-\pi RM} 
   \frac{1}{R^8m_n^8}
  +\frac{1}{2}
   \frac{M^2R^2+3n^3}{R^{10} m_n^{10}}
  +\frac{n^4(-1)^n}{R^{10}m_n^{10}}
 \Bigg], 
\end{align}

\begin{align}
 a(\phi_{y 2})
=& \notag
 \frac{2m_{\mu}}{e}
 \begin{array}{c}
  \includegraphics{fig7.eps}
 \begin{picture}(0,0)(1,1)
   \put(-8,30){$\phi^{(m)}_y$}
   \put(-50,40){$\psi_{2,3}^{(n)}$}
   \put(-50,15){$\psi_{2,3}^{(n)}$}
  \end{picture}
 \end{array}
 \notag \\
\sim&
 \frac{3e^2}{8\pi^3}(MR)^3 R^2 m_\mu
 \bigg[
  m_W S^1_{\phi_y}-
  m_{\mu} S^2_{\phi_y}
  -8(MR)^2 m_W
  \left(S^{3}_{\phi_y}-\frac{4MR}{\pi}S^{4}_{\phi_y}\right)
  \notag \\
&~~~~
  + m_W S^5_{\phi_y}
  - m_{\mu} S^6_{\phi_y}-8 m_W S^7_{\phi_y}
  +2 m_WS^8_{\phi_y}
 \bigg]
\end{align}
where 
\begin{align}
 S^1_{\phi_y}
=&
 \sum_{m,n=1}^{\infty} \dX
 \frac{X^2}{D}
 \left(\frac{1}{R^2m_{n+m}^2}-\frac{1}{R^2m_{n-m}^2}\right)^2
 (-1)^{n+m}\frac{n^2-(MR)^2}{R^2m_n^2}, 
 \nt \\
 S^2_{\phi_y}
=& 
 \sum_{m,n=1}^{\infty}\dX
 \frac{X^2(X-1)}{D}
 \left(\frac{1}{R^2m_{n+m}^2}-\frac{1}{R^2m_{n-m}^2}\right)^2, 
 \nt \\
 S^{3}_{\phi_y}
=&
 \sum_{m,n=1}^{\infty}\dX
 \frac{X^2(-1)^{n+m}}{D}
 \left(
  \frac{1}{R^2m_{n+m}^2}-\frac{1}{R^2m_{n-m}^2}
 \right)
 \left[
  \frac{n(m-n)}{R^4m_{n-m}^4}+\frac{n(n+m)}{2R^4m_{n+m}^4}
 \right], 
 \nt \\
 S^{4}_{\phi_y}
=&
 \sum_{l,m,n=1}^{\infty}\dX
 \frac{X^2}{D}
 \left(
  \frac{1}{R^2m_{n+m}^2}-\frac{1}{R^2m_{n-m}^2}
 \right)
 \frac{(-1)^l(1-(-1)^{l+m+n})l^2mn}{R^4m_l^4[(l+m)^2-n^2][(l-m)^2-n^2]}, 
 \nt \\
 S^5_{\phi_y}
=&
 \sum_{m,n=1}^{\infty}\dX
 \frac{1}{R^4m_n^4}
 \left(
  \frac{2n+m}{R^2m_{n+m}^2}+\frac{2n-m}{R^2m_{n-m}^2}
 \right)^2
 \frac{\{n^2(-1)^{n+m}+(MR)^2\}X^2}{D}, 
 \nt \\
 S^6_{\phi_y}
=&
 \sum_{m,n=1}^{\infty}\dX
 \frac{1}{R^2m_n^2}
 \left(
  \frac{2n+m}{R^2m_{n+m}^2}+\frac{2n-m}{R^2m_{n-m}^2}
 \right)^2
 \frac{X^2(X-1)}{D}, 
 \nt \\
 S^7_{\phi_y}
=&
 \sum_{l,m,n=1}^{\infty}\dX
 \frac{X^2}{D}
 \frac{l}{R^4m_l^4}
 \left(
  \frac{2n+m}{R^2m_{n+m}^2}+\frac{2n-m}{R^2m_{n-m}^2}
 \right)
 (-1)^l
 \frac{\delta_{n+m,l} +\delta_{m+l,n} -\delta_{n+l,m}}{2}, 
 \nt \\
 S^8_{\phi_y}
=&
 \sum_{m,n=1}^{\infty}\dX
 \frac{X^3}{D^2}
 \left[
  \left(\frac{1}{R^2m_{n+m}^2}-\frac{1}{R^2m_{n-m}^2}\right)^2
  +\frac{1}{R^2m_n^2}
  \left(\frac{2n+m}{R^2m_{n+m}^2}+\frac{2n-m}{R^2m_{n-m}^2}\right)
 \right]. 
\end{align}

We have calculated numerically the above mode sums and summarize contributions from neutral current sector.
\begin{align}
 a(h_{y1})
=&
 6.89\times 10^{-9}(Rm_W)^2,
 \\
 a(h_{y2})
=&
 -2.94\times 10^{-8}(Rm_W)^2,
 \\
 a(\gamma_y)
=&
 3.21\times 10^{-7}(Rm_W)^2,
 \\
 a(h_{\mu})
=&
 1.21\times10^{-5}(Rm_W)^2,
 \\
 a(Z_{\mu1})
=&
 6.70\times 10^{-10}(Rm_W)^2,
 \\
 a(Z_{\mu2}) 
=&
 8.56\times 10^{-8}(Rm_W)^2,
 \\
 a(\phi_{y1})
=&
 -9.17\times 10^{-8}(Rm_W)^2,
 \\
 a(\phi_{y2})
=&
 4.14\times 10^{-7}(Rm_W)^2.
\end{align}
Thus, we obtain contributions from neutral sector $a(\text{N.C.})$.

\begin{equation}
\label{a(N.C.)}
 a(\text{N.C.})
=
 1.32\times 10^{-5} (Rm_W)^2. 
\end{equation}

\subsection{Charged current}

The contributions to $g-2$ from the charged current sector are 
due to the following four types of diagrams. 
\begin{align}
\C_1=
 \begin{array}{c}
 \includegraphics{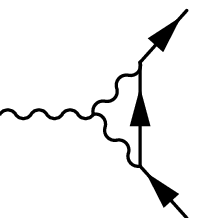}
 \begin{picture}(0,0)(1,1)
 \put(-10,26){$\nu$}
 \put(-48,4){{$W_{\mu}^{(m)}\atop X_{\mu}^{(m)}$}}
 \put(-44,45){{$W_{\mu}^{(m)}\atop X_{\mu}^{(m)}$}}
 \end{picture}
 \end{array}
\sim&
 -\frac{7eg_4^2}{16\pi^2}(RM)^4
 \sum_{m=1}^\infty \frac{1}{m^2(4M^2R^2+m^2)^2}R^2m_{\mu}, 
\end{align}


\begin{align}
\C_{2}
=
 \begin{array}{c}
 \includegraphics{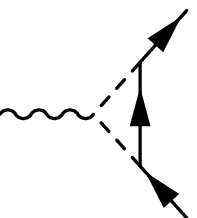}
 \begin{picture}(0,0)(1,1)
 \put(-5,26){$\psi_1^{(n)}$}
 \put(-40,8){{$X_y$}}
 \put(-40,44){{$X_y$}}
 \end{picture}
 \end{array}
~\sim
 -\frac{3eg_4^2}{4\pi^3}M^3R^5
 \sum_{n=1}^\infty \left[\frac{1}{3}m_{\mu}+(-1)^nm_W\right]
 \frac{n^2}{(Rm_n)^8}, 
\end{align}


\begin{align}
\C_{3}
=&
 \begin{array}{c}
 \includegraphics{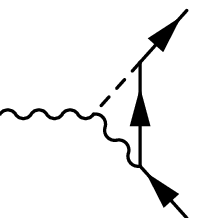}
 \begin{picture}(0,0)(1,1)
 \put(-10,25){$\nu$}
 \put(-50,10){{$W_{\mu}^{(m)}$}}
 \put(-45,45){{$X_y^{(m)}$}}
 \end{picture}
 \end{array}
+\begin{array}{c}
 \includegraphics{fig9.eps}
 \begin{picture}(0,0)(1,1)
 \put(-10,25){$\nu$}
 \put(-50,10){{$X_{\mu}^{(m)}$}}
 \put(-45,45){{$W_y^{(m)}$}}
 \end{picture}
 \end{array}\\
\sim
&
 \frac{3eg_4^2}{2\pi^4} (MR)^4 R^2 m_{\mu}{\rm e}^{\pi RM}
 \sum_{n,m=1}^\infty \frac{(-1)^n n^2}{R^4m_n^4}\frac{1}{R^4m_{n+m}^2m_{n-m}^2}
 \frac{1}{4M^2R^2+m^2},  
\end{align}

\begin{align}
\C_4=&
\begin{array}{c}
\includegraphics{fig8.eps}
\begin{picture}(0,0)(1,1)
 \put(-10,25){$\psi_1^{(n)}$}
 \put(-70,0){\small {$W_{\mu}^{(m)}, X_{\mu}^{(m)}$}}
 \put(-70,45){\small {$W_{\mu}^{(m)}, X_{\mu}^{(m)}$}}
\end{picture}
\end{array}
+
\begin{array}{c}
\includegraphics{fig9.eps}
\begin{picture}(0,0)(1,1)
 \put(-10,25){$\psi_1^{(n)}$}
 \put(-70,0){\small {$W_{\mu}^{(m)}, X_{\mu}^{(m)}$}}
 \put(-70,45){\small {$W_y^{(m)}, X_y^{(m)}$}}
\end{picture}
\end{array}
+
\begin{array}{c}
\includegraphics{fig10.eps}
\begin{picture}(0,0)(1,1)
 \put(-10,25){$\psi_1^{(n)}$}
 \put(-70,0){\small {$W_y^{(m)}, X_y^{(m)}$}}
 \put(-70,45){\small {$W_y^{(m)}, X_y^{(m)}$}}
\end{picture}
\end{array}
\nt\\
\sim&
 -\frac{3eg_4^2}{2\pi^3}(MR)^3(S_1+S_2)R^2m_{\mu}
 +\frac{3eg_4^2}{\pi}(MR)^3S_3R^2m_W
 \nt\\
&
 -\frac{3eg_4^2}{4\pi^3}(MR)^3
 \left(
  4S_{4}+16\frac{MR}{\pi}S_{5}
  +S_6+S_7+S_{8}+4\frac{MR}{\pi}S_9
 \right)R^2m_W
 \nt\\
&
 -\frac{3eg_4^2}{\pi^3}(MR)^3(S_{10}+S_{11})R^2m_W
\end{align}
where
\begin{align}
S_1
=&
\sum_{m,n=1}^{\infty}
\dX\left[
 \frac{M_m^2}{m_n^2}\frac{(1-X)^2(3-2X)}{\D'}
 +\frac{(1-X)^2(3-X)}{\D'}
\right]
\frac{n^2m^2}{(R^4m_{n+m}^2m_{n-m}^2)^2},
\nonumber \\
S_2
=&
\sum_{m,n=1}^{\infty}
\dX\frac{X(1-X)^2}{\D'}\frac{n^2R^2m_n^2}{(R^4m_{n+m}^2m_{n-m}^2)^2},
\nonumber \\ 
S_3
=&
\sum_{m,n=1}^{\infty}
 \frac{n^2m^2}{(R^4m_{n+m}^2m_{n-m}^2)^2}
 \left[
  -\frac{1}{R^2m_n^2-m^2}
  -\dX\frac{R^2m_n^2-2m^2}{R^2m_n^2-m^2}\frac{1}{\D'}
 \right](-1)^{n+m},
\nonumber \\
 S_{4}=&
\sum_{m,n=1}^{\infty}
 \dX \frac{n^2m(-1)^{n+m}}{R^4m_{n+m}^2m_{n-m}^2}
 \left[
  \frac{m-n}{R^4m_{n-m}^4}+\frac{n+m}{2R^4m_{n+m}^4}
 \right]
 \frac{(4X-3)(X-1)}{\D'},
 \nonumber \\
 S_{5}=&
\sum_{l,m,n=1}^{\infty}
 \dX\frac{n^2m(-1)^{n+m}}{R^4m_{n+m}^2m_{n-m}^2}\frac{l}{R^4m_l^4}
 \frac{lm(1-(-1)^{l+m+n})}{[(l+m)^2-n^2][(l-m)^2-n^2]}
 \frac{(4X-3)(X-1)}{\D'},
 \nonumber \\
S_6=&
\sum_{m,n=1}^{\infty}
\dX\frac{(1-X)(6X-5)}{\D'}
\left[
\frac{2(m-n)}{R^4m_{n-m}^4}+\frac{n+m}{R^4m_{n+m}^4}
\right]
\frac{(-1)^{n+m}nm^2}{R^4m_{n+m}^2m_{n-m}^2},
\nonumber \\
S_{7}=&
\sum_{m,n=1}^{\infty}
 \dX
 \left[
  \frac{2(m-n)}{R^4m_{n-m}^4}+\frac{n+m}{R^4m_{n+m}^4}
 \right]
 \frac{(-1)^{n+m}nR^2m_n^2}{R^4m_{n+m}^2m_{n-m}^2}
 \frac{X(1-X)}{\D'},
 \nonumber \\
S_{8}=&
\sum_{m,n=1}^{\infty}
 \dX\frac{n^2m}{R^2m_n^2}(-1)^{n+m}
 \frac{R^2m_n^2+m^2}{R^4m_{n+m}^2m_{n-m}^2}
 \left[
  \frac{m-n}{R^4m_{n-m}^4}+\frac{n+m}{2R^4m_{n+m}^4}
 \right]
 \frac{X-1}{\D'},
 \nonumber \\
S_{9}=&
\sum_{l,m,n=1}^{\infty}
 \dX\frac{X-1}{\D'}\frac{n^2m^2l^2}{R^2m_n^2}
 \frac{(-1)^{n+m}}{R^4m_l^4}\frac{R^2m_n^2+m^2}{R^4m_{n+m}^2m_{n-m}^2}
 \frac{1-(-1)^{l+m+n}}{[(l+m)^2-n^2][(l-m)^2-n^2]},
 \nonumber \\
 S_{10}
=&
\sum_{m,n=1}^{\infty}
 \dX \frac{(1-X)^2(6X-5)}{\D'^2}(-1)^{n+m}
 \frac{n^2m^4}{(R^4m_{n+m}^2m_{n-m}^2)^2},
 \nonumber \\
 S_{11}
=&
\sum_{m,n=1}^{\infty}
 \dX\left[
  \frac{2X(1-X)^2}{\D'^2}+\frac{m^2}{R^2m_n^2}\frac{(1-X)^2}{\D'^2}
 \right]
  (-1)^{n+m}\frac{R^2m_n^2n^2m^2}{(R^4m_{n-m}^2m_{n+m}^2)^2}. 
\end{align}
The denominator $\D'$ is defined as
\begin{equation}
\D'=R^2(Xm_n^2+(1-X)M_m^2).
\end{equation}
We approximated the double and triple mode summations 
 by using $\frac{m_{\mu}}{m_W}\ll 1$.
We get the contributions from the charged sector $a(\text{C.C.})$ similar to those from the neutral sector;
\begin{align}
\label{a(C.C.)}
 a(\text{C.C.})
=&
 \frac{2m_{\mu}}{e}[\C_1+\C_2+\C_3+\C_4]
=
 -5.20\times 10^{-7}(Rm_W)^2. 
\end{align}

\section{Contributions from $\Delta$ and $\Sigma$}

There are several gauge interactions relevant to the calculation of muon $g-2$ up to the 1-loop order. 
In this section, we turn to the $\Delta$ and $\Sigma$ sectors which have not been discussed in the main text. 
Before calculating gauge interactions in detail, we show that the diagrams which contain 
isospin 3/2 components ($\Delta$) and an isospin 1 component with a positive charge ($\Sigma^+$) 
have no contributions to muon $g-2$.

First, the $SU(3)/SU(2)$ sector has an isospin 1/2 and the corresponding gauge bosons couple only 
between multiplets with isospins different by 1/2, $\Psi_3\leftrightarrow \left({\Psi_1\atop \Psi_2 }\right),
\left({\Psi_1\atop \Psi_2 }\right)\leftrightarrow \Sigma, \Sigma\leftrightarrow \Delta$. 
After electroweak symmetry breaking, $\tilde{Y}^0$ (which is the 5th component of the gauge field 
with the quantum number same as Higgs in the SM) takes a VEV $v$, 
quadratic mixing terms appearing in the effective Lagrangian also have the same properties.
Thus, we expect that the diagram which contains $\Delta$ and $\Sigma^+$ in the internal line does not exist and 
$\Delta$ and $\Sigma^+$ have no contributions to muon $g-2$ at 1-loop.

Also gauge fields are mixed with each other via $\tilde Y^0$ VEV $v$ and these effects must be considered.
If we insert the VEV into the $X,\tilde Y^0$ propagator, the VEV causes the transformation of $X,\tilde Y^0$ to $W, Z$ or photon, which in turn  belongs to the adjoint or singlet representations of $SU(2)$ 
and cannot connect fermions with different isospins. On the other hands, the isospin of the internal fermions in the diagrams should be changed in order to construct correct diagrams contributing to the anomalous moment. Thus,
we expect that the diagrams with the insertion of the VEV in the gauge boson propagators have no contribution to muon $g-2$. 

\begin{figure}[hb]
\begin{center}
\includegraphics{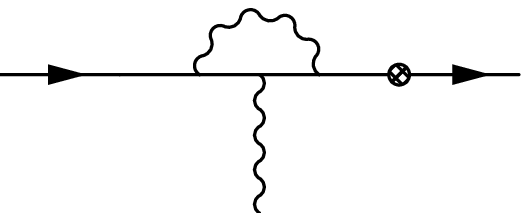}
\begin{picture}(0,0)(1,1)
\put(-25,30){$\mu_R$}
\put(-55,30){$\Psi_2$}
\put(-75,30){$\Sigma^-$}
\put(-95,30){$\Sigma^-$}
\put(-140,30){$\mu_L$}
\put(-80,65){$\tilde Y ^0$}
\end{picture}
\hspace{20mm}
\includegraphics{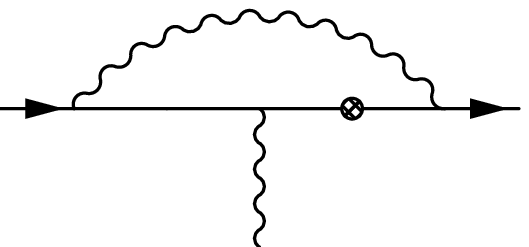}
\begin{picture}(0,0)(1,1)
\put(-20,30){$\mu_R$}
\put(-45,30){$\Psi_2$}
\put(-70,30){$\Sigma^-$}
\put(-105,30){$\Sigma^-$}
\put(-150,30){$\mu_L$}
\put(-80,75){$\tilde Y ^0$}
\end{picture}

\vspace{10mm}
\includegraphics{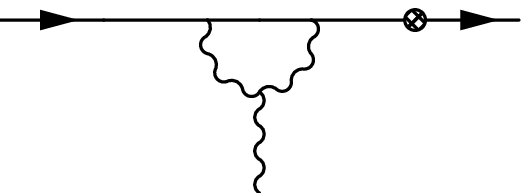}
\begin{picture}(0,0)(1,1)
\put(-20,55){$\mu_R$}
\put(-50,55){$\Psi_2$}
\put(-65,30){$X^-$}
\put(-110,30){$X^-$}
\put(-140,55){$\mu_L$}
\put(-80,55){$\Sigma ^0$}
\end{picture}
\hspace{20mm}
\includegraphics{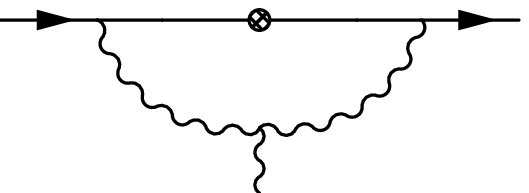}
\begin{picture}(0,0)(1,1)
\put(-20,55){$\mu_R$}
\put(-60,30){$X^-$}
\put(-110,30){$X^-$}
\put(-140,55){$\mu_L$}
\put(-100,55){$\Sigma ^0$}
\put(-60,55){$\Psi_1$}
\end{picture}
\end{center}
\caption{Possible diagrams concerning about $\Sigma$. 
The blob denotes the mass insertion of fermion.}
\label{Possiblediagrams}
\end{figure}

To summarize, the possible diagrams are those where the internal fermion contains 
only $\Sigma^0,\Sigma^-$
and mass insertion into gauge fields should not be considered.
From this observation, we have to consider only four diagrams as shown in figure \ref{Possiblediagrams} 
as the contributions of $\Delta$ and $\Sigma$ to muon $g-2$.

\subsection{Mode functions of fermion}

We impose orbifold boundary conditions to obtain chiral theory as follows:
\begin{equation}
\begin{cases}
\Sigma=\Sigma_L(-,-)+\Sigma_R(+,+),\\
\left({\Psi_1\atop \Psi_2}\right)=\left({\Psi_1\atop \Psi_2}\right)_L(+,+) +
\left({\Psi_1\atop \Psi_2}\right)_R(-,-),\\
\Psi_3={\Psi_3}_L(-,-)+{\Psi_3}_R(+,+)
\end{cases}
\end{equation}
Each 5D fields are expanded in terms of mode functions 
$f_L^{(n)},f_R^{(n)},f_L^{(0)},f_R^{(0)},S_n$
which satisfy eigenvalue equation and boundary conditions.
\begin{equation}
\begin{cases}
\displaystyle
\Sigma(x,y)=\sum_{n=1}^{\infty}[\Sigma_L^{(n)}S_n+\Sigma_R^{(n)}f_R^{(n)}]
+\Sigma_R^{(0)}f_R^{(0)} ,\\
\displaystyle
\left({\Psi_1\atop {\Psi_2}}\right)=\sum_{n=1}^{\infty}
\left[\left({\Psi_1\atop {\Psi_2}}\right)_L^{(n)}f_L^{(n)} +
\left({\Psi_1\atop {\Psi_2}}\right)_R^{(n)}S_n\right]+\left({\nu\atop \mu}\right)_Lf_L^{(0)},\\
\displaystyle
\Psi_3=\sum_{n=1}^{\infty}[{\Psi_3}_L^{(n)}S_n+{\Psi_3}_R^{(n)}f_R^{(n)}] +\mu_Rf_R^{(0)}.
\end{cases}
\end{equation}

\subsection{Gauge interaction}

First, we derive gauge interaction relevant to muon $g-2$.
\begin{align*}
 i\frac{g}{2}\bar{\Psi}\mmot A \Psi
 \supset&
 i\frac{g}{2}\sqrt{2}\tilde Y^{0(m)}_{\mu}
 \Bigg[
  \bar \Sigma^{-(l)}_L\gm \Psi^{(n)}_{2L} (-1)^{n+m+l}I_{sL}^{lmn}
  +\bar \Sigma_R^{-(n)}\gm \Psi_{2R}^{(l)}I_{sL}^{lmn}
\\
&\hspace{20mm}
  +\bar \Sigma ^{-(n)}_L\gm \mu_L(-1)^{n+m}I_{sR}^{0mn}
  +\bar \Sigma_R^{-(0)}\gm \Psi_{2R}^{(n)}I_{sR}^{0mn}
 \Bigg]
\\
&
 +i\frac{g}{2}\sqrt{\frac{3}{2}}\tilde Y_{\mu}^{0(m)}
 \Bigg[
  \bar \Psi_{2L}^{(n)}\gm \Psi_{3L}^{(l)}(-1)^{n+m+l}I_{sL}^{lmn}
  +\bar \Psi_{2R}^{(l)}\gm \Psi_{3R}^{(n)}I_{sL}^{lmn}
\\
&\hspace{20mm}
  +\bar \mu_L\gm \Psi_{3L}^{(n)}(-1)^{n+m}I_{sR}^{0mn}
  +\bar \Psi_{2R}^{(n)}\gm \mu_RI_{sR}^{0mn}
 \Bigg]
\\
&
 +\frac{g}{2}\tilde Y_y^{0(m)}\Bigg[
  -\bar \Sigma_L^{0(n)}\Psi_{1R}^{(l)}I_{c}^{lmn}
  +\bar \Sigma_R^{0(n)}\Psi_{1L}^{(l)}I_{cLR}^{lmn}
  +\bar \Sigma_R^{0(n)}\nu_L(-1)^{(n+m)}I_{cLR}^{nm0}
  +\bar \Sigma_R^{0(0)}\Psi_{1L}^{(n)}I_{cLR}^{nm0}
 \Bigg]
\\
&
 +\frac{g}{2}\frac{1}{\sqrt{2\pi R}}\tilde Y_y^{0(0)}
 \Bigg[
  -\bar \Sigma_L^{0(n)}\Psi_{1R}^{(l)}\delta_{nl}
  +\bar \Sigma_R^{0(n)}\Psi_{1L}^{(l)}I_{LR}^{ln}
\\
&\hspace{20mm}
 +\bar \Sigma_R^{0(n)}\nu_LI_{LR}^{0n}
 +\bar \Sigma_R^{0(0)}\Psi_{1L}^{(n)}(-1)^nI_{LR}^{0n}
 +\bar \Sigma_R^{0(0)}\nu_LI_{LR}^{00}
 \Bigg]
\\
&
 +\frac{g}{2}\sqrt{2}
 \left[\tilde Y_y^0\bar \Sigma ^0\Psi_1\right]_{\Sigma^0\to \Sigma^-,\Psi_1\to \Psi_2}
\\
&
 +\frac{g}{2}\sqrt{\frac{3}{2}}\tilde Y_y^{0(m)}\Bigg[
  -\bar \Psi_{2L}^{(l)}\Psi_{3R}^{(n)}I_{cLR}^{lmn}
  -\bar \mu_L \Psi_{3R}^{(n)}(-1)^{(n+m)}I_{cLR}^{nm0}
  -\bar \Psi_{2L}^{(n)}\mu_RI_{cLR}^{nm0}
  +\bar \Psi^{(n)}_{2R} \Psi_{3L}^{(l)}I_{c}^{lmn}
 \Bigg]
\\
&
 +\frac{g}{2}\frac{1}{\sqrt{2\pi R}}\sqrt{\frac{3}{2}}
 \tilde Y_y^{0(0)}
 \Bigg[
  -\bar \Psi ^{(n)}_{2L} \Psi_{3R}^{(l)}(-1)^{n+l}I_{LR}^{ln}
  -\bar \mu_L \Psi_{3R}^{(n)}I_{LR}^{0n}
\\
&\hspace{20mm}
 -\bar \Psi_{2L}^{(n)}\mu_R(-1)^nI_{LR}^{0n}
 -\bar \mu_L \mu_RI_{LR}^{00}
 +\bar \Psi_{2R}^{(n)}\Psi_{3L}^{(l)}\delta_{nl}
 \Bigg]
\\
&
 +i\frac{g}{\sqrt{2}}
 \left[\tilde Y_{\mu}^0\bar \Sigma^-\gm \Psi_2\right]
 _{\tilde Y^0_{\mu}\to \tilde X_{\mu}^+,\Sigma^-\to \Sigma^+,\Psi_2\to \Psi_1}
 +i\frac{g}{2}
 \left[\tilde Y_{\mu}^0\bar \Sigma^-\gm \Psi_2\right]
 _{\tilde Y^0_{\mu}\to \tilde X_{\mu}^+,\Sigma^-\to \Sigma^0}
\\
&
 +i\frac{g}{2}\sqrt{\frac{3}{2}}
 \left[\tilde Y_{\mu}^0\bar \Psi_2\gm \Psi_3\right]
 _{\tilde Y^0_{\mu}\to \tilde X_{\mu}^+,\Psi_2\to \Psi_1}
 +\frac{g}{\sqrt{2}}
 \left[\tilde Y_y^0\bar \Sigma^0\gamma_5 \Psi_1\right]
 _{\tilde Y^0_y\to \tilde X_y^+,\Sigma^0\to \Sigma^+}
\\
&
 +\frac{g}{2}
 \left[\tilde Y_y^0\bar \Sigma^0\gamma_5 \Psi_1\right]
 _{\tilde Y^0_y\to \tilde X_y^+,\Psi_1\to \Psi_2}
 +\frac{g}{2}\sqrt{\frac{3}{2}}
 \left[\tilde Y_y^0\bar \Psi_2\gamma_5 \Psi_3\right]
 _{\tilde Y^0_y\to \tilde X_y^+,\Psi_2\to \Psi_1}
\end{align*}
where 
$\psi^T=(\Sigma^+,\Sigma^0,\Sigma^-,\Psi_1,\Psi_2,\Psi_3)$ in the first line.
Since $\Sigma_R,\Psi_{1L},\Psi_{2L},\Psi_{3R}$ and $\Sigma_L,\Psi_{1R},\Psi_{2R},\Psi_{3L}$ 
has same $Z_2$ parity, 
we obtain all interactions by replacing fields and numerical factor in the typical one 
.

\subsection{Mass eigenstates of the $SU(2)$ triplet}

In this section,  
we argue mass eigenstate of fermions without $A_y$ VEV. 
A ten dimensional representation is introduced 
to reproduce Standard Model lepton doublet and singlet in the zero mode sector.
However, massless exotic fields appear as $SU(2)$ quartet and triplet, 
{\rm i.e.} $\Delta$ and $\Sigma$, 
which should be removed from the low-energy effective theory 
by introducing brane localized mass terms. 
Fortunately, $\Delta$ has no contributions to muon $g-2$ up to 1-loop order as mentioned above, 
so we concentrate on $\Sigma$ contributions.

The Lagrangian is given by 
\begin{equation}
 \mathcal {L}
=
 \bar \Psi (x,y) (i\partial_M\Gamma^M -M\epsilon(y))\Psi
 +\pi R\delta (y-\pi R)M_3' \frac{1}{\sqrt{2\pi R}}\bar \chi _{3L}(x)
 \Sigma_R(x,y)
\end{equation}
where a four dimensional fermion $\chi_{3L}(x)$ is introduced to couple with $\Sigma_R$. 
We introduce a brane localized mass term at fixed point $y=\pi R$,
since $\Sigma_R^{(0)}$ strongly localizes at $y=\pi R$.
Then we have
\begin{equation}
\label{fermion mass}
\begin{aligned}
\iy \mathcal{L}_{\text{quadratic}}
=&
m_n\left[-\bar {\Psi}^{(n)}_3 {\Psi}^{(n)}_3
    -\left(\bar \Psi_1 \bar \Psi_2\right)^{(n)}
     \left(\begin{array}{c}\bar \Psi_1\\ \bar \Psi_2\end{array}\right)^{(n)}\right]
 -\bar\lambda_L M \lambda_R . 
\end{aligned}
\end{equation}
where
\begin{eqnarray}
\bar \lambda_L M \lambda_R
&=&
 \left(
  \begin{array}{c}
   \bar\chi_{3L} ,\bar\Sigma_L^{(1)} ,\bar\Sigma_L^{(2)},\cdots
  \end{array}
 \right)
 \left[
  \begin{array}{cccc}
   -aM_3' & -f_1 M_3' & -f_2M_3' & \cdots \\
   0 & m_1 & 0 &\cdots \\
   0&0 &m_2& \cdots \\
   \vdots & \vdots& \vdots & \ddots
  \end{array}
 \right]
 \left(
  \begin{array}{c}
   \Sigma_R^{(0)} \\\Sigma_R^{(1)} \\\Sigma_R^{(2)}\\ \vdots
  \end{array}
 \right), \\
a &=& \frac{1}{\sqrt{2}}\sqrt{\frac{\pi RM}{1-\e^{-2\pi RM}}}, \quad f_n=\frac{(-1)^nM_n}{\sqrt{2}m_n} 
\end{eqnarray}
where chiral rotation 
$\left({\Psi_1\atop \Psi_2}\right)\to \e^{-i\frac{\pi}{4}\gamma_5}\left({\Psi_1\atop \Psi_2}\right)$
is performed.

Putting $RM=3.05$, ratio $|f_n/a|$ is smaller than $1/3$, 
which means that it is valid to treat $f_n/a$ as a perturbation.
We obtain fermion mass eigenvalues and corresponding eigenstates (denoted as $\Sigma'$)
up to an order of $\mathcal{O}(f_n/a)$ by performing biunitary transformations as
\begin{equation}
\begin{aligned}
 \text{fermion mass term}
=&
 -m_n(\bar \Psi_1 \Psi_1 +\bar \Psi_2 \Psi_2+\bar \Psi_3\Psi_3)
 -\left[m_n^2+\frac{(f_nM_3'm_n)^2}{m_n^2-a^2M_3'^2}\right]
 \bar \Sigma'^{(n)}\Sigma'^{(n)}
 \\
&
 -a^2M_3'^2\left[1+\sum_n\frac{(f_nM_3')^2}{a^2M_3'^2-m_n^2}\right]\bar \chi'_{3L}\Sigma_R'^{(0)}
 +{\rm h.c.},
\end{aligned}
\end{equation}
where the relations between 
fermion mass eigenstates ($\Sigma'$) and gauge eigenstates ($\Sigma$) can be obtained as follows.
\begin{equation}
\left\{
\begin{aligned}
\chi_{3L}&=\chi_{3L}'-\sum_nC_{0n}\Sigma_L'^{(n)},\\
\Sigma_L^{(n)}&=\Sigma_L'^{(n)}+C_{0n}\chi_{3L}',\\
\Sigma_R^{(0)}&=\Sigma_R'^{(0)}+\sum_n\frac{m_n}{aM_3'}C_{0n}\Sigma_R'^{(n)},\\
\Sigma_R^{(n)}&=\Sigma_R'^{(n)}-\frac{m_n}{aM_3'}C_{0n}\Sigma_R'^{(0)}
\end{aligned}
\right.
\end{equation}
with $C_{0n}=-\frac{f_nM_3'm_n}{a^2M_3'^2-m_n^2}$.

\subsection{Calculation of the $\Sigma$ contributions to muon $g-2$}

\subsubsection{Contributions of $\Sigma'^-$}

\begin{figure}[h]
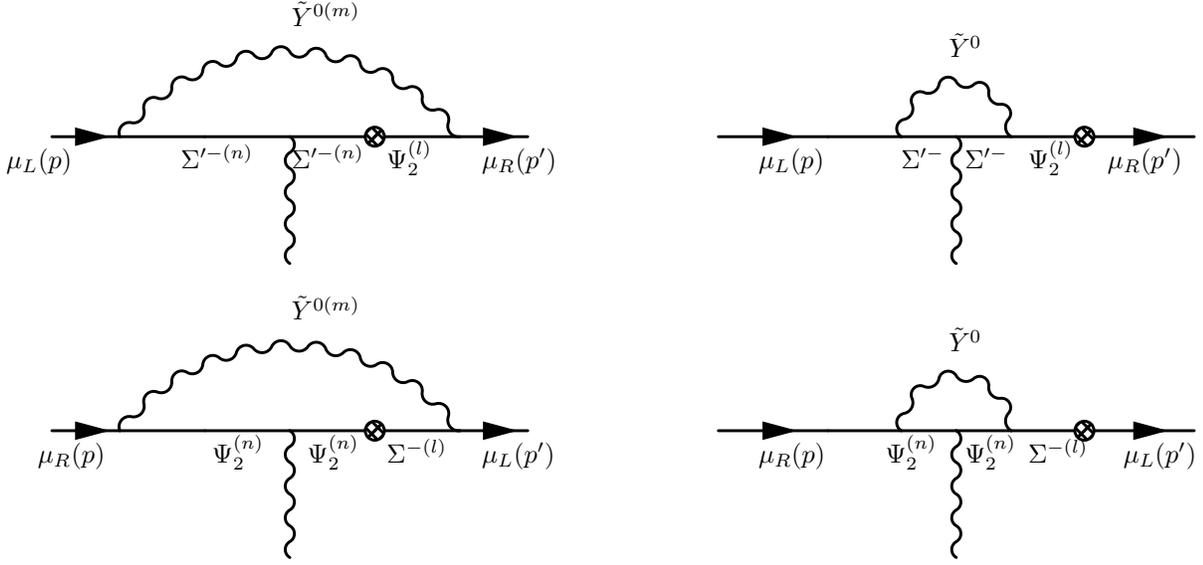

 \begin{center}\small 
 \includegraphics[scale=1.2]{ExVectorInternal.eps}
 \setlength{\unitlength}{1.2pt}
 \begin{picture}(0,0)(1,1)
 \put(-20,30){\footnotesize $\mu_R(p')$}
 \put(-50,30){\footnotesize $\Psi_2^{(l)}$}
 \put(-80,30){\footnotesize $\Sigma'^{-(n)}$}
 \put(-115,30){\footnotesize $\Sigma'^{-(n)}$}
 \put(-170,30){\footnotesize $\mu_L(p)$}
 \put(-80,75){\footnotesize $\tilde Y^{0(m)}$}
 \end{picture}
 \hspace{20mm}
 \includegraphics[scale=1.2]{ExVectorExternal.eps}
 \begin{picture}(0,0)(1,1)
 \put(-30,30){\footnotesize $\mu_R(p')$}
 \put(-55,30){\footnotesize $\Psi_2^{(l)}$}
 \put(-75,30){\footnotesize $\Sigma'^-$}
 \put(-95,30){\footnotesize $\Sigma'^-$}
 \put(-140,30){\footnotesize $\mu_L(p)$}
 \put(-80,65){\footnotesize $\tilde Y ^0$}
 \end{picture}

\vspace{10mm}

 \includegraphics[scale=1.2]{ExVectorInternal.eps}
 \setlength{\unitlength}{1.2pt}
 \begin{picture}(0,0)(1,1)
 \put(-20,30){\footnotesize $\mu_L(p')$}
 \put(-50,30){\footnotesize $\Sigma^{-(l)}$}
 \put(-75,30){\footnotesize $\Psi_2^{(n)}$}
 \put(-105,30){\footnotesize $\Psi_2^{(n)}$}
 \put(-160,30){\footnotesize $\mu_R(p)$}
 \put(-80,75){\footnotesize $\tilde Y^{0(m)}$}
 \end{picture}
 \hspace{20mm}
 \includegraphics[scale=1.2]{ExVectorExternal.eps}
 \begin{picture}(0,0)(1,1)
 \put(-25,30){\footnotesize $\mu_L(p')$}
 \put(-55,30){\footnotesize $\Sigma^{-(l)}$}
 \put(-75,30){\footnotesize $\Psi_2^{(n)}$}
 \put(-100,30){\footnotesize $\Psi_2^{(n)}$}
 \put(-140,30){\footnotesize $\mu_R(p)$}
 \put(-80,65){\footnotesize $\tilde Y ^0$}
 \end{picture}

 \caption{The diagram concerning $\Sigma^{-}$}
\label{fig_neutral}
\end{center}
\end{figure}

In this subsection, 
we investigate the contributions of the charged component of the triplet $\Sigma'^-$ by mass insertion.
We should consider two types of diagrams labeled by the location of mass insertion as in the figure \ref{fig_neutral}.
Chirality flip on external and internal line 
corresponds to replace external momenta with fermion mass
by using fermion equation of motion (EOM) and extracting fermion mass from the internal fermion propagator, respectively.

First, we consider a top-left diagram in figure \ref{fig_neutral}.
The amplitude is calculated in the following. 
\begin{align}
 \mathcal{A}(\Sigma^-)_{\rm int}
=&
 \dk\frac{\sqrt{3}eg^2v}{32\sqrt{2\pi R}}
 \frac{1}{(k^2-M_m^2)[(k+p')^2-m_l^2][(k+p')^2-m_n'^2][(k+p)^2-m_n'^2]}
 \nt\\
&\times
 i\gamma^{\nu}\left[(1+i)V_L'L+(1-i)V_R'R\right]^{l=0}_{n\to l}
 (\mot k+\mot p' +m_l)
 \left[U_L^0(1-i)R+U_R^0(1+i)L\right]^{n\leftrightarrow l}
 \nt\\
&\times
 (\mot k+\mot p' +m_n')
 \gm(\mot k+\mot p+m_n')
 \gamma_\nu 
 \left[V_L(1+i)L+V_R(1-i)R\right]^{l=0}
\end{align}
where $[\cdots]^{l=0}_{n\to l}$ stands for extracting terms linear in $\delta_{l0}$ and replacing index $n$ with $l$.
After straightforward calculations, we have
\begin{align}
 \mathcal{A}(\Sigma^-)_{\rm int}
\sim&
 -\frac{\sqrt{3}eg^2}{16}\frac{M^3R^3}{\pi^2}Rm_W
 \left[3S^{\Sigma'}_{01}-S^{\Sigma'}_{02}+2S^{\Sigma'}_{03}
 +8\frac{MR}{\pi}S^{\Sigma'}_{04}\right]
\end{align}
where
\begin{align}
S^{\Sigma'}_{01}
=&
 \sum_{n,m=1}^{\infty}\dx x(1-x-y)
 \frac{(-1)^{n+m}n^2m^2}{R^8m_{n+m}^4m_{n-m}^4}\frac{1}{\D_{nn}},
 \\
 S^{\Sigma'}_{02}
=&
 \sum_{n,m=1}^{\infty}\dx x(1-x-y)
 \frac{(-1)^{n+m}n^2m^2}{R^8m_{n+m}^2m_{n-m}^2}\frac{R^2m_n'^2}{\D_{nn}^2},
 \\
 S^{\Sigma'}_{03}
=&
 \sum_{n,m=1}^{\infty}\dx x(1-x-y)
 \frac{(-1)^{n+m}((MR)^2-n^2)n^2m^2}{R^8m_{n+m}^4m_{n-m}^4}\frac{m_n'}{m_n}\frac{1}{\D_{nn}^2},
 \\
 S^{\Sigma'}_{04}
=&
 \sum_{n,m,l=1}^{\infty}\dx x(1-x-y)
 \frac{(-1)^{n+m}n^2m^2l^2(1-(-1)^{n+l})(1-\delta_{nl})}{R^8m_{n+m}^2m_{n-m}^2m_{l+m}^2m_{l-m}^2(n^2-l^2)}
 \frac{m_n'}{m_n}\frac{1}{\D _{nl}^2}, \\
\D_{nl}=&R^2[xm_n'^2 +ym_l^2 +(1-x-y)M_m^2]. 
\end{align}
We abbreviated $\int_0^1{\rm d} x\int_0^{1-x}{\rm d}y$ as $\dx$. 

Next, we consider the top-right diagram in figure \ref{fig_neutral}.
Before calculation, 
we provide fermion EOM concerning about zero mode right handed fermion $\mu_R$ 
and KK mode, zero mode left handed doublet $\Psi_{2L}^{(l)}$.
\begin{equation}
\begin{aligned}
 \mot p'\Psi_2^{(l)}=i\frac{gv}{4\sqrt{2\pi R}}\frac{\sqrt{3}}{2}(1-i)
 U_R'^0{}^{l=0}_{n\to l}\mu_R
\end{aligned}
\end{equation}
where the KK mass term is ignored since we are interested in chirality flipping by $A_y$ VEV.
Thus we have
\begin{align}
 \mathcal{A}(\Sigma^-)_{\rm Ext}
=&
 \dk\frac{eg^2}{2}\dx\frac{\sqrt{3}gv}{8\sqrt{2\pi R}}
 \frac
 {I_{sR}^{0mn}[I_{LR}^{0l}I_{sL}^{nml}-I_{LR}^{00}I_{\rm A}^{(nm)}\delta_{l0}](x+y-2)(x+y-1)}
 {[k^2-(x+y)m_n'^2-(1-x-y)M_m^2]^3}P^\mu
\nt\\
\sim &
 -\frac{\sqrt{3}eg^2}{32}\frac{M^3R^3}{\pi^2}Rm_W
 \left[
 \frac{1}{2\pi}S_{05}^{\Sigma'}-4MR S_{06}^{\Sigma'}
 -I_{LR}^{00}S_{07}^{\Sigma'}
 \right]
\end{align}
where
\begin{align}
 S^{\Sigma'}_{05}
=&
 \sum_{n,m,l=1}^{\infty}\dX \frac{X(X-1)(X-2)}{\D}
 \frac{(-1)^{n+m}nml^2}{R^8m_{n+m}^2m_{n-m}^2m_l^4}(\delta_{n+l,m}+\delta_{l+m,n}-\delta_{n+m,l}),
 \\
 S^{\Sigma'}_{06}
=&
 \sum_{n,m,l=1}^{\infty}\dX \frac{X(X-1)(X-2)}{\D}
 \frac{n^2m^2l^2}{R^8m_{n+m}^2m_{n-m}^2m_l^4}
 \frac{(-1)^{n+m}(1-(-1)^{n+m+l})}{[(n+m)^2-l^2][(n-m)^2-l^2]},
 \\
 S^{\Sigma'}_{07}
=&
 \sum_{n,m=1}^{\infty}\dX \frac{X(X-1)(X-2)}{\D}
 \frac{n^2m^2}{R^8m_{n+m}^4m_{n-m}^4}, \\
\D =& R^2[Xm_n'^2+(1-X)M_m^2]. 
\label{D}
\end{align}
$X$ stands for $x+y$.

Similarly, the amplitudes $\mathcal{A'}(\Sigma^-)_{\rm int}
,\mathcal{A'}(\Sigma^-)_{\rm Ext}$ which correspond to 
the bottom-left and the bottom-right diagrams in figure \ref{fig_neutral} are obtained.
\begin{align}
 \mathcal{A'}(\Sigma^-)_{\rm int}
\sim &
  \frac{\sqrt{3}eg^2}{16}\frac{M^3R^3}{\pi^4}Rm_W
  \left[ 3S^{\Sigma'}_{08} + 2S^{\Sigma'}_{09}
 + 8\frac{MR}{\pi}S^{\Sigma'}_{10} - S^{\Sigma'}_{11}
 \right],
 \\
 \mathcal{A'}(\Sigma^-)_{\rm Ext}
\sim &
 \frac{\sqrt{3}eg^2}{64}\frac{M^3R^3}{\pi^4}Rm_W
 \left[
  S^{\Sigma'}_{12} + \frac{8MR}{\pi}m_WS^{\Sigma'}_{13}
 \right]
\end{align}
where
\begin{align}
 S^{\Sigma'}_{08}
=&
 \sum_{n,m=1}^{\infty}\dx x(1-x-y)\frac{(-1)^{n+m}}{\D'_{nn}}
 \frac{n^2m^2}{R^8m_{n+m}^4m_{n-m}^4},
 \\
 S^{\Sigma'}_{09}
=&
 \sum_{n,m=1}^{\infty}\dx x(1-x-y)\frac{(-1)^{n+m}}{\D'^2_{nn}}
 \frac{m_n'}{m_n}\frac{((MR)^2-n^2)n^2m^2}{R^8m_{n+m}^4m_{n-m}^4},
 \\
 S^{\Sigma'}_{10}
=&
 \sum_{n,m,l=1}^{\infty}\dx x(1-x-y)\frac{(-1)^{l+m}}{\D'^2_{nl}}
 \frac{m_l'}{m_l}\frac{n^2m^2l^2(1-(-1)^{n+l})(1-\delta_{nl})}{R^8m_{n+m}^2m_{n-m}^2m_{l+m}^2m_{l-m}^2(n^2-l^2)},
 \\
 S^{\Sigma'}_{11}
=&
 \sum_{n,m=1}^{\infty}\dx x(1-x-y)\frac{(-1)^{n+m}R^2m_n^2}{\D_{nn}'^2}
 \frac{n^2m^2}{R^8m_{n+m}^4m_{n-m}^4},
 \\
 S^{\Sigma'}_{12}
=& 
 \sum_{n,m,l=1}^{\infty}\dX \frac{X(X-1)(X-2)}{\D'}
 \frac{n^2ml}{R^8m_{n+m}^2m_{n-m}^2m_l^3m_n}
 (\delta_{l+n,m}+\delta_{m+n,l}-\delta_{l+m,n}),
 \\
 S^{\Sigma'}_{13}
=&
 \sum_{n,m,l=1}^{\infty}\dX \frac{X(X-1)(X-2)}{\D'}
 \frac{n^2m^2l^2}{R^8m_{n+m}^2m_{n-m}^2m_l^3m_n} 
 \frac{1-(-1)^{l+m+n}}{[(l+m)^2-n^2][(l-m)^2-n^2]}, \\
\D_{nl}' =& R^2[xm_n^2+ym_l'^2+(1-x-y)M_m^2]. 
\label{D'}
\end{align}

\begin{figure}[h]

\vspace{3mm}

\begin{center}
 \includegraphics[scale=1.2]{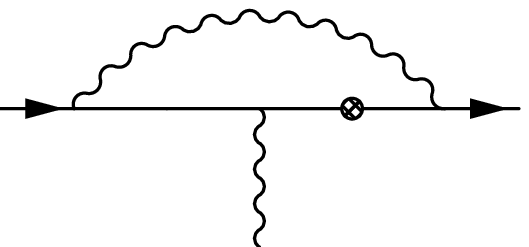}
 \setlength{\unitlength}{1.2pt}
 \begin{picture}(0,0)(1,1)
 \put(-20,30){\footnotesize $\mu_R(p')$}
 \put(-45,30){\footnotesize $\Psi_2^{(l)}$}
 \put(-70,30){\footnotesize $\Sigma^{-(n)}$}
 \put(-105,30){\footnotesize $\Sigma^{-(n)}$}
 \put(-150,30){\footnotesize $\mu_L(p)$}
 \put(-80,75){\footnotesize $\tilde Y ^{0(m)}_y$}
 \end{picture}
 \hspace{20mm}
 \includegraphics[scale=1.2]{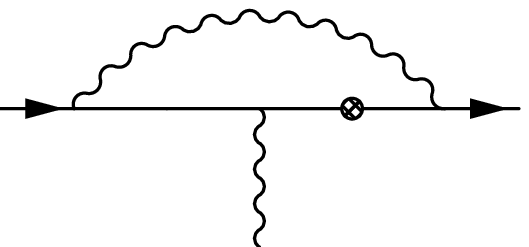}
 \setlength{\unitlength}{1.2pt}
 \begin{picture}(0,0)(1,1)
 \put(-20,30){\footnotesize $\mu_L(p')$}
 \put(-50,30){\footnotesize $\Sigma^{-(l)}$}
 \put(-75,30){\footnotesize $\Psi_2^{(n)}$}
 \put(-105,30){\footnotesize $\Psi_2^{(n)}$}
 \put(-160,30){\footnotesize $\mu_R(p)$}
 \put(-80,75){\footnotesize $\tilde Y^{0(m)}_y$}
 \end{picture}

\vspace{10mm}

 \includegraphics[scale=1.2]{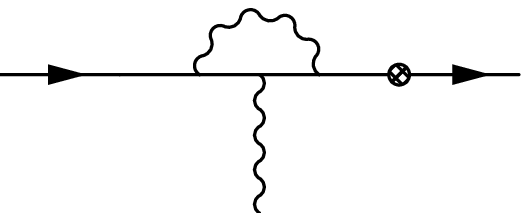}
 \begin{picture}(0,0)(1,1)
 \put(-25,30){\footnotesize $\mu_R(p')$}
 \put(-55,30){\footnotesize $\Psi_2^{(l)}$}
 \put(-75,30){\footnotesize $\Sigma^{-(n)}$}
 \put(-105,30){\footnotesize $\Sigma^{-(n)}$}
 \put(-140,30){\footnotesize $\mu_L(p)$}
 \put(-80,65){\footnotesize $\tilde Y ^{0(m)}_y$}
 \end{picture}
 \hspace{20mm}
 \includegraphics[scale=1.2]{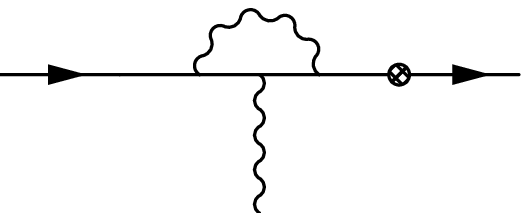}
 \begin{picture}(0,0)(1,1)
 \put(-25,30){\footnotesize $\mu_L(p')$}
 \put(-55,30){\footnotesize $\Sigma^{-(l)}$}
 \put(-75,30){\footnotesize $\Psi_2^{(n)}$}
 \put(-100,30){\footnotesize $\Psi_2^{(n)}$}
 \put(-140,30){\footnotesize $\mu_R(p)$}
 \put(-80,65){\footnotesize $\tilde Y ^{0(m)}_y$}
 \end{picture}
\end{center}
\caption{Scalar exchange diagrams}
\label{fig_scalar_neutral}
\end{figure}

Next, we move on to scalar exchange diagrams.
The diagrams we must consider are shown in figure \ref{fig_scalar_neutral}.
We consider the diagrams in which fermion chirality flips on the internal line. 
The amplitudes 
$\mathcal{B}(\Sigma^-)^{\rm KK,zero}_{\rm int}$,
$\mathcal{B}'(\Sigma^-)^{\rm KK,zero}_{\rm int}$
which correspond to the upper diagrams of figure \ref{fig_scalar_neutral}
are obtained. 
\begin{align}
 \mathcal{B}(\Sigma^-)^{\rm KK}_{\rm int}
 \sim &
 -\frac{\sqrt{3}eg^2}{1024}\frac{M^3R^3}{\pi^4}Rm_W 
 \left[2 S^{\Sigma'}_{14}+\frac{8MR}{\pi}m_W S^{\Sigma'}_{15}
 +S^{\Sigma'}_{16} + \frac{4MR}{\pi}S^{\Sigma'}_{17} + S^{\Sigma'}_{18}
 \right],
 \\
 \mathcal{B}(\Sigma^-)^{\rm zero}_{\rm int}
=&
 \mathcal{B}(\Sigma^-)^{\rm KK}_{\rm int}
 |_{U_{L,R}\to \frac{1}{\sqrt{2\pi R}}U^0_{L,R},U'_{\rm L,R}\to \frac{1}{\sqrt{2\pi R}}U'^0_{L,R}}
 \nt\\
\sim &
 -\frac{\sqrt{3}eg_4^2}{64}\frac{M^3R^3}{\pi^3}R^2m_W
 \left[ 2 S^{\Sigma'}_{19}
 +\frac{8MR}{\pi}S^{\Sigma'}_{20}
 +S^{\Sigma'}_{21}+\frac{4MR}{\pi}S^{\Sigma'}_{22}+2S^{\Sigma'}_{23}
 \right],
 \\
 \mathcal{B'}(\Sigma^-)^{\rm KK}_{\rm int}
\sim &
 -\frac{\sqrt{3}eg^2}{64}\frac{M^3R^3}{\pi^3}Rm_W
 \left[2S^{\Sigma'}_{24}-\frac{8MR}{\pi}m_WS^{\Sigma'}_{25}
  +S^{\Sigma'}_{26} - \frac{4MR}{\pi}S^{\Sigma'}_{27} + 2S^{\Sigma'}_{28}
  \right],
 \\
 \mathcal{B}'(\Sigma^-)^{\rm zero}_{\rm int}
=&
 \mathcal{B'}(\Sigma^-)^{\rm KK}_{\rm int}
 |_{U'_{L,R}\to \frac{1}{\sqrt{2\pi R}}U'^0_{L,R},U_{L,R}\to \frac{1}{\sqrt{2\pi R}}U^0_{L,R}}
 \nt\\
\sim&
  -\frac{\sqrt{3}eg_4^2}{16}\frac{M^3R^3}{\pi^3}Rm_W
 \left[ 2S^{\Sigma'}_{29} + \frac{8MR}{\pi}S^{\Sigma'}_{30}
 - 2 I_{LR}^{00} S^{\Sigma'}_{31} - S^{\Sigma'}_{32}
 - S^{\Sigma'}_{33}
 + \frac{4MR}{\pi}S^{\Sigma'}_{34}
 \right]
\end{align}
where $g_4$ stands for four dimensional gauge coupling:$g_4=\frac{g}{\sqrt{2\pi R}}$.
Summations are defined as follows.
\begin{align}
 S^{\Sigma'}_{14}
=&
 16\sum_{n,m=1}^{\infty}\dx \frac{x(1-\frac{3}{2}x-3y)}{\D_{nn}}
 \frac{(-1)^{n+m}(M^2R^2-n^2)n^2}{R^8m_{n+m}^4m^4_{n-m}},
 \\
 S^{\Sigma'}_{15}
=&
 16\sum_{n,m,l=1}^\infty \dx 
 \frac{x(1-\frac{3}{2}x-3y)}{\D_{nl}}
 \frac{(-1)^{l+m}(1-(-1)^{n+l})n^2l^2}{R^8m_{n+m}^2m_{n-m}^2m_{l+m}^2m_{l-m}^2(n^2-l^2)}
 (1-\delta_{nl})
 ,
 \\
 S^{\Sigma'}_{16}
=& 
 16\sum_{n,m=1}^\infty \dx \frac{x^2}{\D_{nn}}
 \frac{(-1)^{n+m}(M^2R^2-n^2)n^2R^2m_n'^2}{R^8m_{n+m}^4m_{n-m}^4},
 \\
 S^{\Sigma'}_{17}
=&
 16\sum_{n,m,l=1}^\infty\dx 
 \frac{x^2}{\D_{nl}}\frac{(-1)^{l+m}(1-(-1)^{n+l})n^2l^2}{R^8m_{n+m}^2m_{n-m}^2m_{l+m}^2m_{l-m}^2(n^2-l^2)}
 R^2m_n'^2
 (1-\delta_{nl})
 ,
 \\
 S^{\Sigma'}_{18}
=& 
 16\sum_{n,m=1}^\infty \dx \frac{x(x+y)}{\D_{nl}^2}
 (-1)^{n+m}\frac{m_n'}{m_n}
 \frac{n^2R^4m_n^4}{R^8m_{n+m}^4m_{n-m}^4},
 \\
 S^{\Sigma'}_{19}
=&
 \sum_{n,l=1}^\infty \dx
 \frac{x(1-\frac{3}{2}x-3y)}{\D_{nn}^{m=0}}\frac{nl(M^2R^2-n^2)}{R^8m_n^5m_l^3}
 (-1)^l\delta_{nl}
 ,
 \\
 S^{\Sigma'}_{20}
=&
 \sum_{n,l=1}^\infty \dx\frac{x(1-\frac{3}{2}x-3y)}{\D_{nl}^{m=0}}
 \frac{n^2l^2(1-(-1)^{n+l})(1-\delta_{nl})}{R^8m_n^4m_l^4(n^2-l^2)}(-1)^l
 ,
 \\
 S^{\Sigma'}_{21}
=&
 \sum_{n,l=1}^\infty \dx
 \frac{x^2R^2m_n'^2}{(\D_{nn}^{m=0})^2}\frac{nl(M^2R^2-n^2)}{R^8m_n^5m_l^3}
 (-1)^l\delta_{nl}
 ,
 \\
 S^{\Sigma'}_{22}
=&
 \sum_{n,l=1}^\infty \dx 
 \frac{x^2R^2m_n'^2}{(\D_{nl}^{m=0})^2}
 \frac{n^2l^2(1-(-1)^{n+l})(1-\delta_{nl})}{R^8m_n^4m_l^4(n^2-l^2)}(-1)^l
 ,
 \\
 S^{\Sigma'}_{23}
=&
 \sum_{n=1}^\infty \dx
 \frac{x(x+y)R^2m_n'^2}{(\D_{nn}^{m=0})^2}\frac{n^2}{R^6m_n^6},
 \\
 S^{\Sigma'}_{24}
=&
 \sum_{n,m=1}^\infty \dx
 \frac{x(1-\frac{3}{2}x-3y)}{\D_{nn}'}\frac{(-1)^{n+m}(M^2R^2-n^2)n^2}{R^8m_{n+m}^4m_{n-m}^4}
 ,
 \\
 S^{\Sigma'}_{25}
=&
 \sum_{n,m,l=1}^\infty \dx
 \frac{x(1-\frac{3}{2}x-3y)}{\D_{nl}'}\frac{(-1)^{n+m}(1-(-1)^{n+l})n^2l^2}{R^8m_{n+m}^2m_{n-m}^2m_{l+m}^2m_{l-m}^2(n^2-l^2)}(1-\delta_{nl})
 ,
 \\
 S^{\Sigma'}_{26}
=&
 \sum_{n,m=1}^\infty \dx
 \frac{x^2(-1)^{n+m}}{(\D_{nn}')^2}\frac{R^2m_n^2(M^2R^2-n^2)n^2}{R^8m_{n+m}^4m_{n-m}^4}
 ,
 \\
 S^{\Sigma'}_{27}
=&
 \sum_{n,m,l=1}^\infty \dx
 \frac{x^2(-1)^{n+m}}{(\D_{nl}')^2}\frac{(1-(-1)^{n+l})R^2m_n^2n^2l^2}{R^8m_{n+m}^2m_{n-m}^2m_{l+m}^2m_{l-m}^2(n^2-l^2)}(1-\delta_{nl})
 ,
 \\
 S^{\Sigma'}_{28}
=& 
 \sum_{n,m=1}^\infty \dx 
 \frac{x(x+y)R^4m_n^3m_n'}{(\D_{nn}')^2}\frac{(-1)^{n+m}n^2}{R^8m_{n+m}^4m_{n-m}^4}
 ,\\
 S^{\Sigma'}_{29}
=& 
 \sum_{n=1}^\infty \dx \frac{x(1-\frac{3}{2}x-3y)}{\D_{nn}'^{m=0}}\frac{n^2(M^2R^2-n^2)}{R^8m_n^8},
 \\
 S^{\Sigma'}_{30}
=&
 \sum_{n,l=1}^\infty \dx\frac{x(1-\frac{3}{2}x-3y)}{\D_{nl}'^{m=0}}
 \frac{n^2l^2(1-(-1)^{n+l})}{R^8m_n^4m_l^4(l^2-n^2)}(1-\delta_{nl}),
\\
 S^{\Sigma'}_{31}
=&
 \frac{1}{3}
 \sum_{l=1}^\infty \frac{l^2}{R^8m_l'^2m_l^6},
 \\
 S^{\Sigma'}_{32}
=&
 \sum_{n=1}^\infty \dx
 \frac{x^2R^2m_n^2}{(\D_{nn}'^{m=0})^2}\frac{n^2(M^2R^2-n^2)}{R^8m_n^8},
 \\
 S^{\Sigma'}_{33}
=&
 \sum_{n,l=1}^\infty \dx \frac{x^2R^2m_n^2}{(\D_{nl}'^{m=0})^2}\frac{n^2(M^2R^2-n^2)}{R^8m_n^8},
 \\
 S^{\Sigma'}_{34}
=&
 \sum_{n,l=1}^\infty \dx\frac{x^2R^2m_n^2}{\D_{nn}'^{m=0}}
 \frac{n^2l^2(1-(-1)^{n+l})}{R^8m_n^4m_l^4(l^2-n^2)}(1-\delta_{nl}).
\end{align}
${\cal D}_{nl}$ and ${\cal D}'_{nl}$ are given by (\ref{D}) and (\ref{D'})

Next, we consider the diagram in which fermion chirality flips on the external line.
The amplitudes $\mathcal{B}(\Sigma^-)^{\rm KK,zero}_{\rm Ext}$,
$\mathcal{B'}(\Sigma^-)^{\rm KK,zero}_{\rm Ext}$ 
corresponding to bottom diagrams of figure \ref{fig_scalar_neutral} are listed.

\begin{align}
 \mathcal{B}(\Sigma^-)^{\rm KK}_{\rm Ext}
\sim&
 \frac{\sqrt{3}eg^2}{128}\frac{M^3R^3}{\pi^4}Rm_W
 \left[ S^{\Sigma'}_{35} + 2I_{LR}^{00}S^{\Sigma'}_{36}\right],
 \\
 \mathcal{B}(\Sigma^-)^{\rm zero}_{\rm Ext}
=&
 \mathcal{B}(\Sigma^-)^{\rm KK}_{\rm Ext}|_{U_{L,R}\to \frac{1}{\sqrt{2\pi R}}U_{L,R}^0}
 \sim 
 \frac{\sqrt{3}eg_4^2}{384}\frac{M^3R^3}{\pi^3}R^2m_W
 \left[ S^{\Sigma'}_{37} + \frac{4MR}{\pi}S^{\Sigma'}_{38} - I_{LR}^{00}S^{\Sigma'}_{39}
 \right],
 \\
 \mathcal{B'}(\Sigma^-)^{\rm KK}_{\rm Ext}
\sim &
 \frac{\sqrt{3}eg^2}{32}\frac{M^3R^4}{4\pi^4}m_WS^{\Sigma'}_{35}
 \\
 \mathcal{B'}(\Sigma^-)^{\rm zero}_{\rm Ext}
=&
 \mathcal{B'}(\Sigma^-)^{\rm KK}_{\rm Ext}|_{U_{L,R}\to \frac{1}{\sqrt{2\pi R}}U^0_{L,R}}
\sim
 \frac{\sqrt{3}eg_4^2}{384}\frac{M^3R^3}{\pi^3}R^2m_W
 \left[ S^{\Sigma'}_{37} + \frac{4MR}{\pi}S^{\Sigma'}_{40} -S^{\Sigma'}_{41}\right]
\end{align}
where
\begin{align}
 S^{\Sigma'}_{35}
=&
 \sum_{n,m=1}^\infty \dX\frac{X^2(X-1)(-1)^{n+m}}{\D'}
 \\
&
 \times
 \left[
  \frac{n^2(n+m)^2-M^2R^2n(n+m)}{R^8m_{n+m}^6m_{n-m}^2}
  +2\frac{n^2(n-m)^2+M^2R^2n(n-m)}{R^8m_{n+m}^2m_{n-m}^6}
 \right],
 \\
 S^{\Sigma'}_{36}
=&
 \sum_{n,m=1}^\infty \dX\frac{X^2(X-1)}{\D'}\frac{R^2m_n^2n^2}{R^8m_{n+m}^4m_{n-m}^4},
 \\
 S^{\Sigma'}_{37}
=&
 \sum_{n=1}^\infty \frac{(-1)^nn^2(M^2R^2-n^2)}{R^{10}m_n^{10}},
 \\
 S^{\Sigma'}_{38}
=& S^{\Sigma'}_{40} =
 \sum_{n,l=1}^\infty \frac{n^2l^2((-1)^l-(-1)^n)}{R^{10}m_n^6m_l^4(n^2-l^2)}(1-\delta_{nl}),
 \\
 S^{\Sigma'}_{39}
=&
 \sum_{n=1}^\infty \frac{n^2}{R^8m_n^8},
 \\
 S^{\Sigma'}_{41}
=&
 \sum_{l=1}^\infty \frac{1}{R^2m_0^2}\frac{l^2}{R^6m_l^6}, \\
\D' =& R^2[Xm_n^2+(1-X)M_m^2]. 
\end{align}

\subsubsection{Contributions of $\Sigma'^0$}

Now, we calculate the contributions of neutral component of the triplet $\Sigma'^0$.
\begin{figure}[h]
\begin{center}\footnotesize
\includegraphics{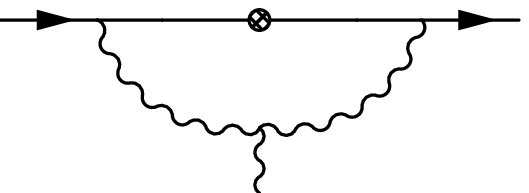}
\begin{picture}(0,0)(1,1)
\put(-20,55){$\mu_R(p')$}
\put(-70,30){$X^{+(m)}_\nu$}
\put(-110,30){$X^{-(m)}_\rho$}
\put(-145,55){$\mu_L(p)$}
\put(-100,55){$\Sigma ^{0(n)}$}
\put(-60,55){$\Psi_1^{(l)}$}
\put(-105,5){$\displaystyle \longrightarrow\atop \displaystyle p-k$}
\put(-70,5){$\displaystyle \longleftarrow\atop \displaystyle k-p'$}
\end{picture}
\hspace{20mm}
\includegraphics{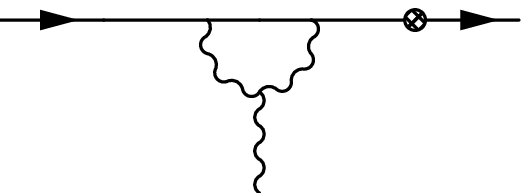}
\begin{picture}(0,0)(1,1)
\put(-20,55){$\mu_R(p')$}
\put(-55,55){$\Psi_2^{(l)}$}
\put(-65,30){$X^{+(m)}_\nu$}
\put(-120,30){$X^{-(m)}_\rho$}
\put(-140,55){$\mu_L(p)$}
\put(-80,55){$\Sigma ^{0(n)}$}
\end{picture}
\end{center}
\caption{Charged current diagrams}
\label{fig_charged}
\end{figure}
The diagrams which must be considered are shown in figure \ref{fig_charged}.
Note that the self interactions
\footnote{
Feynman rules for self interactions are derived in figure \ref{selfinteraction}.
}
contain interaction between gauge bosons,  gauge boson and scalar boson, scalar bosons.
So, these diagrams stand for all possible diagrams 
by replacing internal gauge fields $X^{-(m)}_\mu$ by charged scalar boson $X^{-(m)}_y$.

Let us first consider the left diagram in figure \ref{fig_charged} and its amplitude $\mathcal{A}(\Sigma^0)_{\rm int}$.
The subscript ``int" means the mass insertion in the internal fermion line. 
The superscript \lq\lq $\mu\mu$\rq\rq,\lq\lq $\mu y$\rq\rq,\lq\lq $yy$\rq\rq
are understood as a sort of gauge self-interactions.

\begin{align}
 \mathcal{A}(\Sigma^0)^{\mu\mu}_{\rm int}
\sim &
 \frac{3\sqrt{3}eg^2}{128} \frac{M^3 R^3}{\pi^4} Rm_W
 \left[ -3S^{\Sigma'}_{42} + S^{\Sigma'}_{43} + \frac{4MR}{\pi}S^{\Sigma'}_{44}
 \right]
\end{align}
where
\begin{align}
 S^{\Sigma'}_{42}
=&
 \sum_{n,m=1}^\infty \dx\frac{x^2}{\C^m_{nn}}\frac{n^2m^2}{R^8m_{n+m}^4m_{n-m}^4},
 \\
 S^{\Sigma'}_{43}
=&
 \sum_{n,m=1}^\infty \dx\frac{x^2R^2m'_nm_n}{(\C_{nn}^m)^2}
 \frac{(-1)^{n+m}(M^2R^2-n^2)n^2m^2}{R^{10}m_{n+m}^4m_{n-m}^4m_n^2},
 \\
 S^{\Sigma'}_{44}
=&
 \sum_{n,m,l=1}^\infty \dx\frac{x^2}{(\C_{nl}^m)^2}\frac{m_n'}{m_n}
 \frac{(-1)^{n+m}(1-(-1)^{n+l})n^2m^2l^2(1-\delta_{nl})}
      {R^8m_{n+m}^2m_{n-m}^2m_{l+m}^2m_{l-m}^2(l^2-n^2)}, \\
\C_{nl}^m =& R^2[xM_m^2+ym_n'^2+(1-x-y)m_l^2].       
\label{C}
\end{align}

Next, we consider the diagrams where one of the internal gauge boson is replaced by the charged scalar.
There are two possible diagrams. 
One of them is that the charged gauge boson on the right side $X^-_\rho$ is replaced, 
and the other one is that the charged gauge boson on the left side $X^-_\rho$ is replaced.
In this case, we pick up terms linear in KK mass $m$ since Yukawa coupling flips fermion chirality.
\begin{align}
 \mathcal{A}(\Sigma^0)^{\mu y}_{\rm int}
\sim &
 -\frac{\sqrt{3}eg^2}{128}\frac{M^3R^3}{\pi^4}Rm_W
 \left[S^{\Sigma'}_{45}-\frac{4MR}{\pi}S^{\Sigma'}_{46}+S^{\Sigma'}_{47}\right],
 \\
 \mathcal{A}(\Sigma^0)^{y\mu}_{\rm int}
\sim &
 \frac{\sqrt{3}eg^2}{128}\frac{M^3R^3}{\pi^4}Rm_W
 \left[ S^{\Sigma'}_{48} + S^{\Sigma'}_{49} + \frac{4MR}{\pi}S^{\Sigma'}_{50}
 \right], 
 \\
 \mathcal{A}(\Sigma^0)^{yy}_{\rm int,KK}
\sim &
 \frac{\sqrt{3}eg^2}{64}\frac{M^3R^3}{\pi^4}Rm_W
 \left[ 2S^{\Sigma'}_{51} + \frac{8MR}{\pi}S^{\Sigma'}_{52} - S^{\Sigma'}_{53}
 \right],
 \\
 \mathcal{A}(\Sigma^0)^{yy}_{\rm int,zero}
=&
 \mathcal{A}(\Sigma^0)^{yy}_{\rm int,KK}
 |_{U_{L,R}\to \frac{1}{\sqrt{2\pi R}}U_{L,R}^0,U'_{L,R}\to \frac{1}{\sqrt{2\pi R}}U_{L,R}'^0}
 \nt\\
\sim &
 \frac{\sqrt{3}eg_4^2}{192}\frac{M^3R^3}{\pi^3}R^2m_W
  \left[ 6S^{\Sigma'}_{54} + 24\frac{MR}{\pi}S^{\Sigma'}_{55}
  + I_{LR}^{00}S^{\Sigma'}_{56} + 3S^{\Sigma'}_{57}
  \right]
\end{align}
where we mean by ``KK" and ``zero" that internal gauge bosons are nonzero KK modes and zero mode, respectively 
and
\begin{align}
 S^{\Sigma'}_{45}
=&
 \sum_{n,m=1}^\infty \dx\frac{x^2}{(\C_{nn}^m)^2}\frac{m_n'}{m_n}
 \frac{(-1)^{n+m}n^2m^2(M^2R^2-n^2)}{R^8m_{n+m}^4m_{n-m}^4},
 \\
 S^{\Sigma'}_{46}
=&
 \sum_{n,m,l=1}^\infty \dx\frac{x^2}{(\C_{nl}^m)^2}\frac{m_n'}{m_n}
 \frac{(-1)^{n+m}(1-(-1)^{n+l})n^2m^2l^2(1-\delta_{nl})}
      {R^8m_{n+m}^2m_{n-m}^2m_{l+m}^2m_{l-m}^2(l^2-n^2)}
 =S^{\Sigma'}_{44},
 \\
 S^{\Sigma'}_{47}
=&
 \sum_{n,m=1}^\infty \dx\frac{x^2}{(\C_{nn}^m)^2}
 \frac{(-1)^{n+m}R^2m_n^2n^2m^2}{R^8m_{n+m}^4m_{n-m}^4},
 \\
 S^{\Sigma'}_{48}
=&
 \sum_{n,m=1}^\infty \dx\frac{x^2}{(\C_{nn}^m)^2}
 \frac{(-1)^{n+m}R^2m_nm_n'n^2m^2}{R^8m_{n+m}^4m_{n-m}^4},
 \\
 S^{\Sigma'}_{49}
=&
 \sum_{n,m=1}^\infty \dx\frac{x^2}{(\C_{nn}^m)^2}
 \frac{(-1)^{n+m}n^2m^2(M^2R^2-n^2)}{R^8m_{n+m}^4m_{n-m}^4},
 \\
 S^{\Sigma'}_{50}
=&
 \sum_{n,m,l=1}^\infty \dx\frac{x^2}{(\C_{nl}^m)^2}\frac{m_l}{m_n}
 \frac{(-1)^{n+m}(1-(-1)^{n+l})n^2m^2l^2(1-\delta_{nl})}
      {R^8m_{n+m}^2m_{n-m}^2m_{l+m}^2m_{l-m}^2(n^2-l^2)},
  \\
 S^{\Sigma'}_{51}
=&
 \sum_{n,m=1}^\infty \dx 
 \frac{x(\frac{3}{2}x-1)}{\C'^m_{nn}}\frac{(-1)^{n+m}n^2(M^2R^2-n^2)}{R^8m_{n+m}^4m_{n-m}^4},
 \\
 S^{\Sigma'}_{52}
=&
 \sum_{n,m,l=1}^\infty \dx\frac{x(\frac{3}{2}x-1)}{\C_{nl}'^m}
 \frac{(-1)^{m+l}(1-(-1)^{n+l})n^2l^2}
      {R^8m^2_{n+m}m^2_{n-m}m^2_{l+m}m^2_{l-m}(n^2-l^2)}(1-\delta_{nl}),
 \\
 S^{\Sigma'}_{53}
=&
 \sum_{n,m=1}^\infty \dx\frac{x(x-1)}{(\C_{nn}'^m)^2}
 \frac{(-1)^{n+m}n^2R^4m_n^3m_n'}{R^8m_{n+m}^4m_{n-m}^4},
 \\
 S^{\Sigma'}_{54}
=&
 \sum_{n=1}^\infty \dx\frac{x(\frac{3}{2}x-1)}{\C_{nn}^{m=0}}
 \frac{(-1)^nn^2(M^2R^2-n^2)}{R^8m_n^8},
 \\
 S^{\Sigma'}_{55}
=&
 \sum_{n,l=1}^\infty \dx\frac{x(\frac{3}{2}x-1)}{\C_{nl}^{m=0}}
 \frac{n^2l^2((-1)^l-(-1)^n)}{R^8m_n^4m_l^4(n^-l^2)}(1-\delta_{nl}),
 \\
 S^{\Sigma'}_{56}
=&
 3S^{\Sigma'}_{31},
 \\
 S^{\Sigma'}_{57}
=&
 \sum_{n=1}^\infty \dx\frac{x(x-1)}{(\C_{nn}^{m=0})^2}
 \frac{(-1)^nn^2}{R^4m_n^4}\frac{m_n'}{m_n}, \\
\C_{nl}'^m =& R^2[xM_m^2+ym_n^2+(1-x-y)m_l'^2].
\label{C'} 
\end{align}
${\cal C}_{nl}^m$ is given by (\ref{C}).

Next, we consider the right diagram in figure \ref{fig_charged} and
the corresponding amplitudes $\mathcal{A}(\Sigma^0)_{\rm Ext}$ are calculated. 
The subscript ``ext" means the mass insertion in the external fermion line. 
\begin{align}
 \mathcal{A}(\Sigma^0)^{\mu\mu}_{\rm Ext}
\sim &
 -\frac{\sqrt{3}eg_4^2}{128}\frac{M^3R^3}{\pi^3}R^2m_W
 \left[S^{\Sigma'}_{58}+2I_{LR}^{00}S^{\Sigma'}_{59}\right],
 \\
 \mathcal{A}(\Sigma^0)^{y\mu}_{\rm Ext}
=&
 \mathcal{A}(\Sigma^0)^{\mu y}_{\rm Ext}
=
 0,
 \\
 \mathcal{A}(\Sigma^0)^{yy}_{\rm Ext,KK}
 \sim&
 -\frac{\sqrt{3}eg_4^2}{384}\frac{M^3R^3}{\pi^3}R^2m_W
 \left[S^{\Sigma'}_{60}+\frac{4MR}{\pi}S^{\Sigma'}_{61}
 -I_{LR}^{00}S^{\Sigma'}_{62}
 \right],
 \\
 \mathcal{A}(\Sigma^0)^{yy}_{\rm Ext,zero}
\sim & 
 -\frac{\sqrt{3}eg^2}{128}\frac{M^3R^3}{\pi^4}Rm_W
 \left[S^{\Sigma'}_{63}-\frac{8MR}{\pi}S^{\Sigma'}_{64}
 +2I_{LR}^{00}S^{\Sigma'}_{65}\right]
\end{align}
where
\begin{align} 
 S^{\Sigma'}_{58}
=&
 \sum_{n,m=1}^\infty \dX\frac{X^2(X-1)}{\C'^m_n}
 \frac{(-1)^{n+m}}{R^4m_{n+m}^2m_{n-m}^2}
 \nt\\
&
 \times
 \left[\frac{n^2(n+m)^2-M^2R^2n(n+m)}{R^4m_{n+m}^4}+2\frac{n^2(n-m)^2-M^2R^2n(n-m)}{R^4m_{n-m}^4}\right],
 \\
 S^{\Sigma'}_{59}
=&
 \sum_{n,m=1}^\infty \dX\frac{X^2(X-1)}{\C_n'^m}\frac{n^2R^2m_n^2}{R^8m_{n+m}^4m_{n-m}^4},
 \\
 S^{\Sigma'}_{60}
=&
 \sum_{n=1}^\infty \frac{(-1)^nn^2(M^2R^2-n^2)}{R^{10}m_n^8m_n'^2},
 \\
 S^{\Sigma'}_{61}
=&
 \sum_{n,l=1}^\infty \frac{n^2l^2((-1)^l-(-1)^n)}{R^{10}m_n^4m_l^4m_n'^2(n^2-l^2)}(1-\delta_{nl}),
 \\
 S^{\Sigma'}_{62}
=&
 \sum_{n=1}^\infty \frac{n^2}{R^8m_n^6m_n'^2}
 =S^{\Sigma'}_{56}=3S^{\Sigma'}_{31},
 \\
 S^{\Sigma'}_{63} 
=&
 \sum_{n,m=1}^\infty \dX\frac{X^2(X+\frac{1}{2})}{\C_n'^m}
 \frac{(-1)^{n+m}nm}{R^4m_{n+m}^2m_{n-m}^2}
 \left[\frac{2(n-m)^2}{R^4m_{n-m}^4}-\frac{(n+m)^2}{R^4m_{n+m}^4}\right],
 \\
 S^{\Sigma'}_{64}
=&
 \sum_{n,m,l=1}^\infty \dX
 \frac{X^2(X+\frac{1}{2})}{\C_n'^m}
 \frac{n^2m^2l^2}{R^8m_{n+m}^2m_{n-m}^2m_l^4}
 \frac{(-1)^{n+m}(1-(-1)^{l+m+n})}{[(n+m)^2-l^2][(n-m)^2-l^2]},
 \\
 S^{\Sigma'}_{65}
=&
 \sum_{n,m=1}^\infty \dX\frac{X^2(X+\frac{1}{2})}{\C_n'^m}
 \frac{n^2m^2}{R^8m_{n+m}^4m_{n-m}^4},\\
\C_n'^m =& R^2[XM_m^2+(1-X)m_n'^2].  
\end{align}

If the brane localized mass is taken to be infinity, $M_3'\to \infty$, 
the coefficient $C_{0n}$ and mass spectra are close to
\begin{equation}
\left\{
\begin{aligned}
&
 \lim_{M_3'\to \infty}\frac{m_n}{aM_3'}C_{0n}
=0,\\
&
 \lim_{M_3'\to \infty} m_n'
=
 m_n\sqrt{1-\frac{f_n^2}{a^2}}.
\end{aligned}
\right.
\end{equation}
We also approximate the mode sum by using $\e^{\pi RM}\gg 1$.

Zero mode left handed (doublet) and right handed (singlet) fermions 
remain massless ($A_y$ vev is considered by mass insertion)
and zero mode triplet fermion has heavy mass $\sim\mathcal{O}(M_3')$
by introducing brane localized mass term.
Thus in the case of $n=0$, $m'_0$ has a mass of infinity and 
$m_{0}$ is equal to 0.

We notice an infrared divergence in ($S^{\Sigma'}_{41}$).
In our calculation, the base is not mass eigenstate and 
it is regarded as a contribution of massless mode.
Our purpose is to calculate contributions of nonzero KK modes 
and we therefore ignore it.

\subsection{Results of numerical calculation}

Finally, we obtain the magnitude of $g-2$ $a(\Sigma')$ from the contributions of $\Sigma'$.
Some parameters used in the following calculation are given.
\begin{equation*}
\begin{cases}
MR=3.05,\\
\displaystyle
I_{LR}^{00}=\frac{m_{\mu}}{m_W}=1.3\times 10^{-3},\\
\displaystyle
g^2=2\pi Rg_4^2=2\pi R \frac{1}{\sin^2 \theta_W}(\underbrace{\sin\theta_W g_4}_e)^2=\frac{8\pi R}{3}e^2. 
\end{cases}
\end{equation*}
The final result is found as
\begin{equation}
a(\Sigma')=\frac{2m_\mu}{e}\mathcal{A}(\Sigma')
=5.75 \times 10^{-8}(Rm_W)^2.
\end{equation}
It is very small compared with the contribution to $g-2$ 
from the doublet and singlet contributions (\ref{a(N.C.)}) and (\ref{a(C.C.)}). 
Thus, we can safely neglect the contributions from $\Sigma'$.

\subsection{Feynman Rules}
\label{Mod Feynman}

In this section, we summarize Feynman rules used in our calculations. 

\subsubsection{Propagator}
\paragraph{Gauge boson('t Hooft-Feynman gauge)}

\begin{equation}
 \begin{array}{c}
 \includegraphics[scale=1.2]{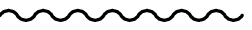}
 \setlength{\unitlength}{1.2pt}
 \begin{picture}(0,0)(1,1)
 \put(0,7){$A_{\nu}^{(m)}$}
 \put(-90,7){$A_{\mu}^{(n)}$}
 \put(-45,10){$\displaystyle p \atop \displaystyle \to$}
 \end{picture}
 \end{array}
=
 \frac{\eta_{\mu\nu}}{p^2-M_n^2}\delta_{nm}.
\end{equation}

\paragraph{Scalar('t Hooft-Feynman gauge)}

\begin{equation}
 \begin{array}{c}
 \includegraphics[scale=1.2]{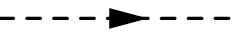}
 \setlength{\unitlength}{1.2pt}
 \begin{picture}(0,0)(1,1)
 \put(0,8){$A_{y}^{(m)}$}
 \put(-90,8){$A_{y}^{(n)}$}
 \put(-45,10){$\displaystyle p \atop \displaystyle \to$}
 \end{picture}
 \end{array}
=
 -\frac{1}{p^2-M_n^2}\delta_{nm}.
\end{equation}

\paragraph{Fermion}

\begin{eqnarray}
 \begin{array}{c}
 \includegraphics[scale=1.2]{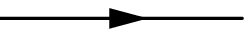}
 \setlength{\unitlength}{1.2pt}
 \begin{picture}(0,0)(1,1)
 \put(0,8){$\mu$}
 \put(-80,8){$\mu$}
 \put(-45,13){$\displaystyle p \atop \displaystyle \to$}
 \end{picture}
 \end{array}
&=&
 -\frac{1}{\mot p}, \\
 \begin{array}{c}
 \includegraphics[scale=1.2]{fermion.eps}
 \setlength{\unitlength}{1.2pt}
 \begin{picture}(0,0)(1,1)
 \put(0,8){$\Psi_{1,2}^{(m)}$}
 \put(-80,8){$\Psi_{1,2}^{(n)}$}
 \put(-45,13){$\displaystyle p \atop \displaystyle \to$}
 \end{picture}
 \end{array}
&=&
 -\frac{1}{\mot p-m_n}\delta_{nm}~~(n,m\neq 0), \\
 \begin{array}{c}
 \includegraphics[scale=1.2]{fermion.eps}
 \setlength{\unitlength}{1.2pt}
 \begin{picture}(0,0)(1,1)
 \put(0,7){$\Sigma'^{(m)}$}
 \put(-80,7){$\Sigma'^{(n)}$}
 \put(-45,12){$\displaystyle p \atop \displaystyle \to$}
 \end{picture}
 \end{array}
&=&
 \frac{-1}{\mot p-m_n'}\delta_{nm}~~(n,m\neq 0), \\
 \begin{array}{c}
 \includegraphics[scale=1.2]{fermion.eps}
 \setlength{\unitlength}{1.2pt}
 \begin{picture}(0,0)(1,1)
 \put(0,7){$\Sigma'^{(0)}$}
 \put(-80,7){$\Sigma'^{(0)}$}
 \put(-45,12){$\displaystyle p \atop \displaystyle \to$}
 \end{picture}
 \end{array}
&=&
 -\frac{1}{\mot k -aM_3'\sqrt{1+\sum_n\frac{(f_nM_3')^2}{a^2M_3'^2-m_n^2}}},
\end{eqnarray}
where $M_n=\frac{n}{R},m_n=\sqrt{M^2+M_n^2},
m'_n=m_n\sqrt{1+\frac{(f_nM_3')^2}{m_n^2-a^2M_3'^2}}$.
$\Sigma$ means both $\Sigma^{-}$ and  $\Sigma^0$.

\subsubsection{Vertex}

\begin{align}
\begin{array}{c}
\includegraphics[scale=1.2]{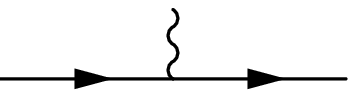}
 \setlength{\unitlength}{1.2pt}
\begin{picture}(0,0)(1,1)
\put (-100,10){$\Psi_2^{(l)}$}
\put (-45,15){$\tilde Y^0_{\mu}$}
\put (-15,10){$\bar\Sigma'^{-(n)}$}
\end{picture}
\end{array}
&=
 \frac{g}{2}\gm(1+i\gamma_5)\left[V_LL+V_RR\right],
 \\
\begin{array}{c}
\includegraphics[scale=1.2]{intVector.eps}
 \setlength{\unitlength}{1.2pt}
\begin{picture}(0,0)(1,1)
\put (-100,10){$\Psi_3$}
\put (-45,15){$\tilde Y^0_{\mu}$}
\put (-15,10){$\bar \Psi_2$}
\end{picture}
\end{array}
&=
 \frac{g}{2}\frac{\sqrt{3}}{2}(1+i\gamma_5)\gm\left[V'_LL+V'_RR\right],
 \\
\begin{array}{c}
\includegraphics[scale=1.2]{intVector.eps}
 \setlength{\unitlength}{1.2pt}
\begin{picture}(0,0)(1,1)
\put (-100,10){$\Psi_2$}
\put (-45,15){$\tilde X^+_{\mu}$}
\put (-15,10){$\bar \Sigma'^0$}
\end{picture}
\end{array}
&=
 \frac{g}{2}\frac{\sqrt{2}}{2}(1+i\gamma_5)\gm\left[V_LL+V_RR\right],
 \\
\begin{array}{c}
\includegraphics[scale=1.2]{intVector.eps}
 \setlength{\unitlength}{1.2pt}
\begin{picture}(0,0)(1,1)
\put (-100,10){$\Psi_3$}
\put (-45,15){$\tilde X^+_{\mu}$}
\put (-15,10){$\bar \Psi_1$}
\end{picture}
\end{array}
&=
 \frac{g}{2}\frac{\sqrt{3}}{2}(1+i\gamma_5)\gm\left[V'_LL+V'_RR\right],
 \\
\begin{array}{c}
\includegraphics[scale=1.2]{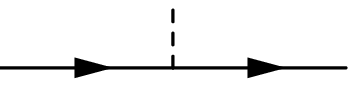}
 \setlength{\unitlength}{1.2pt}
\begin{picture}(0,0)(1,1)
\put (-100,10){$\Psi_1$}
\put (-45,15){$\tilde Y^{0(m)}_y$}
\put (-15,10){$\bar\Sigma'^0$}
\end{picture}
\end{array}
&=
 -i\frac{g}{2}\frac{\sqrt{2}}{2}(1+i\gamma_5)\left[U_LL+U_RR\right],
 \\
\begin{array}{c}
\includegraphics[scale=1.2]{intVector5.eps}
 \setlength{\unitlength}{1.2pt}
\begin{picture}(0,0)(1,1)
\put (-100,10){$\Psi_1$}
\put (-45,15){$\tilde Y^{0(0)}_y$}
\put (-15,10){$\bar\Sigma'^0$}
\end{picture}
\end{array}
&=
 -i\frac{g}{2\sqrt{2\pi R}}\frac{\sqrt{2}}{2}(1+i\gamma_5)\left[U^0_LL+U^0_RR\right],
 \\
\begin{array}{c}
\includegraphics[scale=1.2]{intVector5.eps}
 \setlength{\unitlength}{1.2pt}
\begin{picture}(0,0)(1,1)
\put (-100,10){$\Psi_2$}
\put (-45,15){$\tilde Y^{0(m)}_y$}
\put (-15,10){$\bar\Sigma'^-$}
\end{picture}
\end{array}
&=
 -i\frac{g}{2}(1+i\gamma_5)\left[U_LL+U_RR\right],
 \\
\begin{array}{c}
\includegraphics[scale=1.2]{intVector5.eps}
 \setlength{\unitlength}{1.2pt}
\begin{picture}(0,0)(1,1)
\put (-100,10){$\Psi_2$}
\put (-45,15){$\tilde Y^{0(0)}_y$}
\put (-15,10){$\bar\Sigma'^-$}
\end{picture}
\end{array}
&=
 -i\frac{g}{2\sqrt{2\pi R}}(1+i\gamma_5)\left[U^0_LL+U^0_RR\right],
 \\
\begin{array}{c}
\includegraphics[scale=1.2]{intVector5.eps}
 \setlength{\unitlength}{1.2pt}
\begin{picture}(0,0)(1,1)
\put (-100,10){$\Psi_3$}
\put (-45,15){$\tilde Y^{0(m)}_y$}
\put (-15,10){$\bar \Psi_2$}
\end{picture}
\end{array}
&=
 -i\frac{g}{2}\frac{\sqrt{3}}{2}(1+i\gamma_5)\left[U'_LL+U'_RR\right],
 \\
\begin{array}{c}
\includegraphics[scale=1.2]{./intVector5.eps}
 \setlength{\unitlength}{1.2pt}
\begin{picture}(0,0)(1,1)
\put (-100,10){$\Psi_3$}
\put (-45,15){$\tilde Y^{0(0)}_y$}
\put (-15,10){$\bar \Psi_2$}
\end{picture}
\end{array}
&=
 -i\frac{g}{2\sqrt{2\pi R}}\frac{\sqrt{3}}{2}(1+i\gamma_5)\left[{U'}^0_LL+{U'}^0_RR\right],
 \\
\begin{array}{c}
\includegraphics[scale=1.2]{./intVector5.eps}
 \setlength{\unitlength}{1.2pt}
\begin{picture}(0,0)(1,1)
\put (-100,10){$\Psi_2$}
\put (-45,15){$\tilde X^{+(m)}_y$}
\put (-15,10){$\bar \Sigma'^0$}
\end{picture}
\end{array}
&=
 -i\frac{g}{2}\frac{\sqrt{2}}{2}(1+i\gamma_5)\left[U_LL+U_RR\right],
 \\
\begin{array}{c}
\includegraphics[scale=1.2]{./intVector5.eps}
 \setlength{\unitlength}{1.2pt}
\begin{picture}(0,0)(1,1)
\put (-100,10){$\Psi_2$}
\put (-45,15){$\tilde X^{+(0)}_y$}
\put (-15,10){$\bar \Sigma'^0$}
\end{picture}
\end{array}
&=
 -i\frac{g}{2\sqrt{2\pi R}}\frac{\sqrt{2}}{2}(1+i\gamma_5)\left[U^0_LL+U^0_RR\right],
 \\
\begin{array}{c}
\includegraphics[scale=1.2]{./intVector5.eps}
 \setlength{\unitlength}{1.2pt}
\begin{picture}(0,0)(1,1)
\put (-100,10){$\Psi_3$}
\put (-45,15){$\tilde X^{+(m)}_y$}
\put (-15,10){$\bar \Psi_1$}
\end{picture}
\end{array}
&=
 -i\frac{g}{2}\frac{\sqrt{3}}{2}(1+i\gamma_5)\left[U'_LL+U'_RR\right],
 \\
\begin{array}{c}
\includegraphics[scale=1.2]{./intVector5.eps}
 \setlength{\unitlength}{1.2pt}
\begin{picture}(0,0)(1,1)
\put (-100,10){$\Psi_3$}
\put (-45,15){$\tilde X^{+(0)}_y$}
\put (-15,10){$\bar \Psi_1$}
\end{picture}
\end{array}
&=
 -i\frac{g}{2\sqrt{2\pi R}}\frac{\sqrt{3}}{2}(1+i\gamma_5)\left[{U'}^0_LL+{U'}^0_RR\right],
\\
\begin{array}{c}
\includegraphics[scale=1.2]{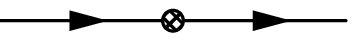}
 \setlength{\unitlength}{1.2pt}
\begin{picture}(0,0)(1,1)
\put (-100,10){$\Psi_1$}
\put (-15,10){$\bar \Sigma'^0$}
\end{picture}
\end{array}
&=
 -i\frac{gv}{4\sqrt{2\pi R}}\frac{\sqrt{2}}{2}(1+i\gamma_5)\left[U^0_LL+U^0_RR\right],
 \\
\begin{array}{c}
\includegraphics[scale=1.2]{./intVEV.eps}
 \setlength{\unitlength}{1.2pt}
\begin{picture}(0,0)(1,1)
\put (-100,10){$\Psi_2$}
\put (-15,10){$\bar \Sigma'^-$}
\end{picture}
\end{array}
&=
 -i\frac{gv}{4\sqrt{2\pi R}}(1+i\gamma_5)\left[U^0_LL+U^0_RR\right],
 \\
\begin{array}{c}
\includegraphics[scale=1.2]{./intVEV.eps}
 \setlength{\unitlength}{1.2pt}
\begin{picture}(0,0)(1,1)
\put (-100,10){$\Psi_3$}
\put (-15,10){$\bar \Psi_2$}
\end{picture}
\end{array}
&=
 -i\frac{gv}{4\sqrt{2\pi R}}\frac{\sqrt{3}}{2}(1+i\gamma_5)
 \left[U'^0_LL+U'^0_RR\right].
\end{align}
where vertex functions are defined as follows. 
\begin{align}
&\left\{
\begin{aligned}
 V_L&=(-1)^{n+m+l}I_{sL}^{(nml)}
      +(-1)^{n+m}I_{sR}^{(0mn)}\delta_{l0},
 \\
 V_R&= I_{sL}^{(lmn)}+\frac{m_n}{aM_3'}C_{0n}I_{sR}^{(0ml)},
\end{aligned}
\right.
\\
&\left\{
\begin{aligned}
 V'_L&=(-1)^{n+m+l}I_{sL}^{(lmn)}
      +(-1)^{m+l}I_{sR}^{(0ml)}\delta_{n0},
 \\
 V'_R&= I_{sL}^{(nml)}+I_{sR}^{(0mn)}\delta _{l0},
\end{aligned}
\right.
\\
&\left\{
\begin{aligned}
 U_L&=I_{cLR}^{(lmn)}+(-1)^{(n+m)}I_{cLR}^{nm0}\delta _{l0}
      +C_{0n}\frac{m_n}{aM_3'}I_{cLR}^{lm0},
 \\
 U_R&= -I_{c}^{(lmn)},
\end{aligned}
\right.
\\
&\left\{
\begin{aligned}
 U_L^0&=I_{LR}^{(ln)}+I_{LR}^{0n}\delta_{l0}+(-1)^l\frac{m_n}{aM_3'}C_{0n}I_{LR}^{0l}
        +\frac{m_n}{aM_3'}C_{0n}\delta_{l0}I_{LR}^{00},
 \\
 U_R^0&= -\delta_{nl},
\end{aligned}
\right.
\\
&\left\{
\begin{aligned}
 U'_L&=I_{c}^{(lmn)},
 \\
 U'_R&=-I_{cLR}^{(nml)}-(-1)^{m+l}I_{cLR}^{lm0}\delta_{n0}
       -(-1)^{n+m}(-1)^{(n+m)}I_{cLR}^{nm0}\delta_{l0},
\end{aligned}
\right.
\\
&\left\{
\begin{aligned}
 U_L'^0&=\delta_{nl},
 \\
 U_R'^0&=-(-1)^{n+l}I_{LR}^{ln}-I_{LR}^{0l}\delta_{n0}-(-1)^nI_{LR}^{0n}\delta_{l0}-I_{LR}^{00}\delta_{n0}\delta_{l0}.
\end{aligned}
\right.
\end{align}
The range of KK index $n,m,l$ is understood to be taken as $1,2,3\cdots$, 
and $\delta_{n0},\cdots$ stands for $n=0$.
And $C_{0n}=-\frac{f_nM_3'm_n}{a^2M_3'^2-m_n^2}$.


\end{document}